\documentclass[iop]{emulateapj}  

\usepackage{color}

\newcommand\kms{\ifmmode{\rm km\thinspace s^{-1}}\else km\thinspace s$^{-1}$\fi}
\newcommand\epic{EPIC~219394517}
\newcommand\ktwo{{\it K2\/}}
\newcommand\kepler{{\it Kepler\/}}

\shortauthors{Torres et al.}
\shorttitle{Eclipsing binary in Ruprecht~147}

\begin{document} 

\submitted{Accepted for publication in The Astrophysical Journal}

\title{Eclipsing binaries in the open cluster Ruprecht~147. I: \epic}

\author{
Guillermo Torres\altaffilmark{1},
Jason L.\ Curtis\altaffilmark{2},
Andrew Vanderburg\altaffilmark{3,4},
Adam L.\ Kraus\altaffilmark{3}, and
Aaron Rizzuto\altaffilmark{3}
}

\altaffiltext{1}{Harvard-Smithsonian Center for Astrophysics, 60
  Garden St., Cambridge, MA 02138, USA; gtorres@cfa.harvard.edu}

\altaffiltext{2}{Department of Astronomy, Columbia University, New
  York, NY 10027, USA}

\altaffiltext{3}{Department of Astronomy, The University of Texas at
  Austin, Austin, TX 78712, USA}

\altaffiltext{4}{NASA Sagan Fellow}

\begin{abstract}

Eclipsing binaries in star clusters offer more stringent tests of
stellar evolution theory than field binaries because models must not
only match the binary properties, but also the radiative properties of
all other cluster members at a single chemical composition and a
single age. Here we report new spectroscopic observations of the G
type, detached eclipsing binary \epic\ in the open cluster
Ruprecht~147 (${\rm [Fe/H]} = +0.10$), which was observed in late 2015
by the \ktwo\ mission. A joint analysis of our radial-velocity
measurements and the \ktwo\ light curve shows the 6.5 day orbit to be
very nearly circular.  We derive highly precise masses of
$1.0782^{+0.0019}_{-0.0019}$~$\mathcal{M}_{\sun}^{\rm N}$ and
$1.0661^{+0.0027}_{-0.0021}$~$\mathcal{M}_{\sun}^{\rm N}$, radii of
$1.055 \pm 0.011$~$\mathcal{R}_{\sun}^{\rm N}$ and $1.042 \pm
0.012$~$\mathcal{R}_{\sun}^{\rm N}$, and effective temperatures of
5930~$\pm$~100~K and 5880~$\pm$~100~K for the primary and secondary,
respectively. The distance we infer, $283^{+18}_{-16}$~pc, corresponds
to a parallax in good agreement with the {\it Gaia}/DR2 value for the
star. Current stellar evolution models from the MIST and PARSEC series
match the above physical properties very well at ages of 2.48 and
2.65~Gyr. Isochrones for these same ages and the measured composition,
along with our reddening estimate for \epic, also show generally good
agreement with the optical and near-infrared color-magnitude diagrams
of the cluster, which can be constructed with no free parameters as
the distances of all member stars are known from {\it Gaia}.

\end{abstract}


\section{Introduction}
\label{sec:introduction}

Star clusters have long been used to test our knowledge of stellar
physics, by comparing observations in the color-magnitude diagram
against models of stellar evolution. This allows inferences to be made
about the age of the cluster and its distance, often aided by
knowledge of the chemical abundance that can be derived
spectroscopically from one or more of its brighter members.
Constraints of a very different nature on stellar theory may be
obtained from suitable detached double-lined eclipsing binaries by
measuring their component masses and radii \citep[see,
  e.g.,][]{Andersen:1991, Torres:2010a}, two of the most fundamental
stellar properties. Accurate values for these properties can usually
be derived from purely geometric and dynamical principles, with no
further assumptions.

When eclipsing binaries are found in a cluster of known metallicity
the constraints become much stronger and the test more valuable, as
models must then match not only the masses and radii of both binary
components at the measured composition, but also the radiative
properties of all other cluster members at the same age as inferred
for the binary. While some three dozen eclipsing binaries have been
studied photometrically and spectroscopically in clusters and young
associations, not all binaries are suitable (detached) for this sort
of test, or have parent populations sufficiently well characterized
(well defined color-magnitude diagrams, spectroscopically known
metallicity), or have had their properties measured well enough.  Some
examples of systems with sufficiently precise measurements permitting
such tests include those published by \cite{Kaluzny:2006},
\cite{Meibom:2009}, \cite{Brogaard:2011, Brogaard:2012},
\cite{Sandquist:2016}, \cite{Brewer:2016}, and others.

Here we present an analysis of a relatively bright ($V = 11.4$)
detached eclipsing binary in the open cluster Ruprecht~147 (NGC~6774),
designated \epic\ (also TYC~6296-96-1 and 2MASS~J19152465-1651222).
Ruprecht~147 lies on the ecliptic, and was observed by \ktwo, the
re-purposed \kepler\ mission, in the final months of 2015. \epic\ was
identified as a cluster member and a double-lined spectroscopic binary
by \cite{Curtis:2013}, and listed as entry CWW~64 in their catalog.
These authors presented the first detailed study of this neglected
cluster, establishing it to be middle-aged ($\sim$3~Gyr), of slightly
super-solar composition (${\rm [Fe/H]} = +0.10$), and distant some
300~pc from the Sun, making it the oldest nearby cluster (see their
Figure~1). What makes Ruprecht~147 special is that it contains no
fewer than \emph{five} eclipsing binaries \citep{Curtis:2016},
offering an unprecedented opportunity for testing stellar evolution
models over a wide range of masses at a single age and composition.
Furthermore, the recently published second data release (DR2) from the
{\it Gaia} consortium \citep{Gaia:2016, Gaia:2018} now provides highly
accurate parallaxes for other cluster members that greatly simplifies
the study of the color-magnitude diagram of Ruprecht~147.

This paper begins our study of the eclipsing systems in the cluster
with \epic, a 6.527~day detached eclipsing binary composed of similar
G-type main-sequence stars. We describe our preparation of the
\ktwo\ data for analysis in Section~\ref{sec:photometry}, our imaging
observations to examine the vicinity of \epic\ in
Section~\ref{sec:imaging}, and our spectroscopic monitoring to derive
radial velocities for both components in
Section~\ref{sec:spectroscopy}. The joint analysis of the \ktwo\ light
curve and the velocities is presented in Section~\ref{sec:lightcurve}.
We then proceed to derive the absolute dimensions of the components
(masses, radii, etc.)\ in Section~\ref{sec:dimensions}. Rotation is
studied in Section~\ref{sec:rotation}, along with the signs of
activity (spots) that we see manifested in \epic\ as quasi-periodic
variability in the light curve. Then in Section~\ref{sec:models} we
compare two sets of current stellar evolution models against our
measurements of the physical properties (mass, radius, temperature),
and illustrate the good agreement (with no adjustable parameters) of
the same best-fit models with brightness measurements for other
cluster members in the optical and near-infrared color-magnitude
diagrams. We conclude in Section~\ref{sec:discussion} with our final
remarks.

\section{\ktwo\ photometry}
\label{sec:photometry}

The Ruprecht~147 cluster was observed by \ktwo\ during Campaign~7 of
the mission, for a period of 75 days (2015 October to December).  The
pixel-level data were downloaded from the Mikulski Archive for Space
Telescopes (MAST)\footnote{\url https://archive.stsci.edu/index.html}
after calibration by the Kepler pipeline \citep{Quintana:2010,
  Jenkins:2010}, and the light curves for \epic\ were then extracted.
We performed a first-pass correction for systematic errors introduced
by \kepler's unstable pointing following \citet{Vanderburg:2014}. The
typical pointing drift of about one pixel over the 6-hour interval
between thruster firings can occasionally be as large as two pixels.
Because Ruprecht~147 is located near the Galactic plane ($b \approx
-13\arcdeg$) in a crowded region of sky, there are several fainter
stars blended with the target star (see below).  To keep the ``third
light'' contribution from these nearby stars constant as the pointing
changes, we extracted the raw photometry from circular moving
apertures with a range of different radii, instead of the stationary
apertures typically used for \ktwo\ data analysis in the procedures of
\citet{Vanderburg:2014}. We chose to use the one giving the lowest
scatter, with a radius of 3.87 pixels, corresponding to 15\farcs4.
Partial pixels were handled by calculating the exact overlap of the
area of the pixel with the circular aperture. We inspected this
first-pass light curve of \epic\ and identified the binary star
eclipses, of which there are 12 primary minima and 13 secondary minima
recorded over the 75-day time span. We then re-derived the
\ktwo\ systematics correction by performing a least-squares fit
simultaneously to the binary star eclipses, the \ktwo\ roll
systematics, and long-term stellar/instrumental variability, following
the procedure described by \citet{Vanderburg:2016}. For the purposes
of deriving the \ktwo\ systematics correction we modeled the long-term
variability with a basis spline in time with nodes spaced every 0.75
days, the \kepler\ roll systematics with cubic splines in
``arclength'' (the position of \kepler's pointing along a
one-dimensional arc), and the binary star eclipses with a
\citet{Mandel:2002} transit model (allowing separate shapes for the
primary and secondary eclipses).  Even though the \citet{Mandel:2002}
model does not perfectly describe binary star eclipses, it gives a
good enough approximation to the light curve shape to yield
high-quality corrections for systematics.

\epic\ shows variability at the 1\% peak-to-peak level that we
interpret as being due mainly to rotational modulation. This obscures
subtler effects such as Doppler beaming, reflection, and ellipsoidal
variations due to the binary orbit. To deal with this, we removed all
long-timescale signals from the light curve (including the rotational
variability, and any phase-dependent variations in the binary) by
dividing the corrected photometry by the best-fit basis spline from
our fit to the variability, systematics, and eclipses.  This rectified
light curve is the one used for our analysis below.  The scatter out
of eclipse in this residual light curve is about 95 parts per million
per 30-minute cadence.

\section{Imaging observations}
\label{sec:imaging}

\begin{figure}
\epsscale{1.15}
\plotone{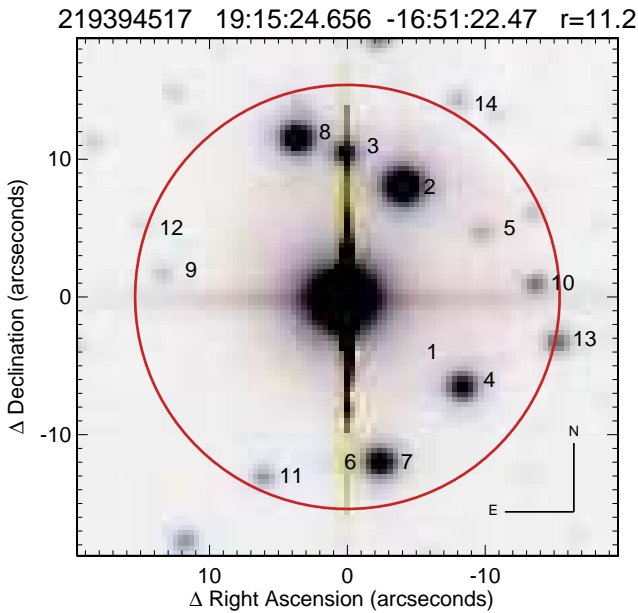}
\figcaption{CFHT $r'$-band image of the field of \epic, with the
  15\farcs4 photometric aperture used to extract the \ktwo\ photometry
  indicated with a circle. Nearby companions are numbered as in
  Table~\ref{tab:CFHT}. Companions \#1 and \#6 are too faint to
  influence the light curve analysis.\label{fig:CFHT}}
\end{figure}

The 15\farcs4-radius photometric aperture contains a number of fainter
stars around \epic\ that will affect the light curve and potentially
bias the parameters derived from it.  Figure~\ref{fig:CFHT} shows a
Sloan $r'$-band image from observations obtained in 2008 by
\cite{Curtis:2013}, which were made with the MegaCam instrument
\citep{Hora:1994} on the Canada-France-Hawaii Telescope (CFHT).
Measurements of the brightness of all numbered companions within the
aperture in the Sloan $g'r'i'$ filters are listed in
Table~\ref{tab:CFHT}, along with their positions relative to our
target.  Many of these companions have entries in the {\it Gaia}/DR2
\citep{Gaia:2018} (see table), though none appear to be cluster
members based on their proper motions (and parallaxes, when
statistically significant).  Also included in the table are $J$- and
$K$-band measurements based on UKIRT/WFCAM imaging
\citep{Curtis:2016}. We use this information below in
Section~\ref{sec:lightcurve} to make a quantitative estimate of the
flux contamination.

\setlength{\tabcolsep}{4.5pt}
\begin{deluxetable*}{rccccccccccccc}
\tablewidth{0pt}
\tablecaption{Neighbors of \epic\ \label{tab:CFHT}}
\tablehead{
\colhead{} &
\colhead{R.A.} &
\colhead{Dec.} &
\colhead{P.A.} &
\colhead{$\rho$} & 
\colhead{$g'$} &
\colhead{$r'$} &
\colhead{$i'$} &
\colhead{$\sigma(gri)$} & 
\colhead{$J$} &
\colhead{$K$} &
\colhead{$\sigma(JK)$} &
\colhead{$G$} &
\colhead{$\pi_{\rm DR2}$}
\\
\colhead{\#} &
\colhead{(J2000)} &
\colhead{(J2000)} &
\colhead{(deg)} &
\colhead{($\arcsec$)} &
\colhead{(mag)} &
\colhead{(mag)} &
\colhead{(mag)} &
\colhead{(mag)} &
\colhead{(mag)} &
\colhead{(mag)} &
\colhead{(mag)} &
\colhead{(mag)} &
\colhead{(mas)}
}
\startdata
 1 & 19:15:24.355 & $-$16:51:26.84 & 225.3 & \phn6.3 & $\cdots$ & $\cdots$ & $\cdots$ & $\cdots$ & 18.57 & 17.90 & 0.07 &    14.60  & 0.129~$\pm$~0.034 \\
 2 & 19:15:24.368 & $-$16:51:14.30 & 333.0 & \phn9.3 & 15.13 & 14.53 & 14.26 & 0.02 & 13.16 & 12.69 & 0.02 &                \nodata &    \nodata \\
 3 & 19:15:24.642 & $-$16:51:11.87 & 358.7 & 10.6 & 17.36 & 16.87 & 16.64 & 0.02 & 15.82 & 15.41 & 0.02 &                    17.06  & 0.300~$\pm$~0.105 \\
 4 & 19:15:24.080 & $-$16:51:28.83 & 232.8 & 10.7 & 17.07 & 16.37 & 16.21 & 0.02 & 15.09 & 14.52 & 0.02 &                    16.70  & 0.089~$\pm$~0.106 \\
 5 & 19:15:23.981 & $-$16:51:17.62 & 296.7 & 11.3 & 20.88 & 19.91 & 20.10 & 0.10 & 18.67 & 17.97 & 0.07 &                    20.17  &    \nodata \\
 6 & 19:15:24.753 & $-$16:51:34.47 & 173.5 & 12.0 & $\cdots$ & $\cdots$ & $\cdots$ & $\cdots$ & 18.65 & 17.74 & 0.06 &      \nodata &    \nodata \\
 7 & 19:15:24.478 & $-$16:51:34.32 & 192.4 & 12.1 & 16.75 & 16.18 & 16.01 & 0.02 & 14.95 & 14.54 & 0.02 &                    16.23  & 0.712~$\pm$~0.069 \\
 8 & 19:15:24.874 & $-$16:51:10.78 & \phn14.8 & 12.2 & 16.37 & 15.62 & 15.33 & 0.02 & 14.06 & 13.47 & 0.02 &                 15.70  & 0.219~$\pm$~0.053 \\
 9 & 19:15:25.523 & $-$16:51:20.62 & \phn81.3 & 13.1 & 23.18 & 22.08 & 21.38 & 0.91 & 18.20 & 17.40 & 0.04 &                 20.80  & 2.973~$\pm$~2.185 \\
10 & 19:15:23.725 & $-$16:51:21.48 & 274.4 & 14.0 & 19.46 & 18.96 & 18.66 & 0.06 & 17.66 & 17.21 & 0.04 &                    18.88  & 0.039~$\pm$~0.385 \\
11 & 19:15:25.038 & $-$16:51:35.38 & 157.1 & 14.1 & 21.01 & 19.80 & 19.14 & 0.02 & 17.36 & 16.52 & 0.02 &                    19.62  &$-$0.048~$\pm$~0.584\phs \\
12 & 19:15:25.618 & $-$16:51:17.01 & \phn68.2 & 15.4 & 24.50 & 23.67 & 21.95 & 0.10 & 18.54 & 17.72 & 0.06 &                \nodata &    \nodata \\
13 & 19:15:23.619 & $-$16:51:25.61 & 258.3 & 15.9 & 19.04 & 18.58 & 18.34 & 0.02 & 17.24 & 16.81 & 0.03 &                   \nodata &    \nodata \\
14 & 19:15:24.097 & $-$16:51:08.02 & 330.9 & 16.8 & 22.02 & 20.83 & 19.98 & 0.02 & 17.96 & 17.02 & 0.03  &                   20.37  &    \nodata
\enddata
\tablecomments{Coordinates derive from the astrometric solutions of the CFHT images \citep[see][]{Curtis:2013}.}
\end{deluxetable*}
\setlength{\tabcolsep}{6pt}

To search for even closer companions along the line of sight to
\epic\ we obtained natural guide star adaptive optics (AO) imaging and
non-redundant aperture-mask interferometry \citep[NRM;
  see][]{Tuthill:2006, Tuthill:2010}.  These observations were
conducted in the $K'$ band ($\lambda = 2.124~\mu$m) on 2016 May 11
with the NIRC2 instrument on the 10\,m Keck~II telescope.  The data
were acquired, reduced, and analyzed following the procedures of
\citet{Kraus:2016}. Table~\ref{tab:AOlimits} lists the $K'$ detection
limits from AO as a function of angular separation from the eclipsing
binary ranging from 150~mas to 2000~mas, expressed as a magnitude
difference relative to the target. They are based on a sequence of $5
\times 10$~s exposures of the object. The limiting apparent magnitude
for companions at the widest separations is $K^\prime \approx
18.2$~mag.  For the NRM observation we used the star EPIC~219511354
(2MASS~J19153533$-$1634117) as a calibrator.  We obtained and analyzed
8 interferograms exposed for 20 seconds each, following observing and
analysis procedures described by \cite{Kraus:2008, Kraus:2011,
  Kraus:2016} and \cite{Rizzuto:2016}.  No companions were detected
from these observations within the limits given in
Tables~\ref{tab:AOlimits} and \ref{tab:apmask}. For reference, these
contrast limits would correspond to companion masses of
0.40~$M_{\sun}$ at 20~mas (5.7~au), 0.17~$M_{\sun}$ at 150~mas
(43~au), and 0.08~$M_{\sun}$ at 500~mas (140~au), if physically bound
to the target.

\begin{deluxetable*}{ccccccccccccc}
\tabletypesize{\footnotesize}
\tablewidth{0pt}
\tablecaption{Keck/NIRC2 $K'$ Imaging Detection Limits (mag) \label{tab:AOlimits}}
\tablehead{
\colhead{MJD} &
\colhead{Number of} &
\colhead{Total} & 
 \multicolumn{10}{c}{Contrast Limit ($\Delta K'$ in mag) at Projected Separation ($\rho$ in mas) } \\
\colhead{} &
\colhead{Frames} &
\colhead{Exposure (s)} & 
\colhead{150} &
\colhead{200} &
\colhead{250} &
\colhead{300} & 
\colhead{400} &
\colhead{500} &
\colhead{700} &
\colhead{1000} &
\colhead{1500} &
\colhead{2000} 
}
\startdata
57520.5 & 5 &  50.00 &  5.6 &  6.6 &  7.1 &  7.5 &  7.9 &  8.1 &  8.2 &  8.4 &  8.4 &  8.4 
\enddata
\tablecomments{No objects were detected within these limits.}
\end{deluxetable*}

\begin{deluxetable*}{lccccccc}
\tabletypesize{\footnotesize}
\tablewidth{0pt}
\tablecaption{Keck/NIRC2 Aperture-Masking Interferometry Detection Limits (mag) \label{tab:apmask}}
\tablehead{
\colhead{Confidence} &
\colhead{MJD} &
\multicolumn{6}{c}{Contrast Limit ($\Delta K'$ in mag) at Projected Separation ($\rho$ in mas) } \\ 
\colhead{Interval} &
\colhead{} &
\colhead{10--20} &
\colhead{20--40} &
\colhead{40--80} &
\colhead{80--160} & 
\colhead{160--240} &
\colhead{240--320}
}
\startdata              
99.9\%      & 57520.5 &  0.79 &  3.83 &  4.73 &  4.51 &  3.95 &  2.86\\ 
99\% only   & 57520.5 &  1.23 &  4.05 &  4.93 &  4.69 &  4.15 &  3.11  
\enddata
\tablecomments{No objects were detected within these limits.}
\end{deluxetable*}

\section{Spectroscopic observations}
\label{sec:spectroscopy}

We monitored \epic\ spectroscopically at the Harvard-Smithsonian
Center for Astrophysics (CfA) between 2016 May and 2017 November with
the Tillinghast Reflector Echelle Spectrograph
\citep[TRES;][]{Furesz:2008}, a bench-mounted fiber-fed echelle
instrument attached to the 1.5m Tillinghast reflector at the Fred
L.\ Whipple Observatory on Mount Hopkins (Arizona, USA). A total of 20
spectra were obtained at a resolving power of $R \approx 44,000$
covering the wavelength region 3800--9100~\AA\ in 51 orders. The
signal-to-noise ratios in the order containing the \ion{Mg}{1}~b
triplet ($\sim$5187~\AA) range from 28 to 39 per resolution element of
6.8~\kms.

Double lines are clearly visible in all our spectra, and there is no
evidence of light from additional stars. Radial velocities for the
components were measured using the two-dimensional cross-correlation
algorithm TODCOR \citep{Zucker:1994}, with templates (one for each
star in the binary) taken from a large library of synthetic spectra
based on model atmospheres by R.\ L.\ Kurucz
\citep[see][]{Nordstrom:1994, Latham:2002}. For these determinations
we used only the echelle order centered on the \ion{Mg}{1}~b triplet,
given that previous experience with similar material shows it contains
most of the information on the velocity \citep[e.g.,][]{Geller:2015,
  Vanderburg:2016}, and because our template library is restricted to
a relatively narrow spectral region centered at this wavelength. The
optimal template for each component was found by running grids of
cross-correlations over a wide range of effective temperatures
($T_{\rm eff}$) and rotational broadenings ($v \sin i$) following
\cite{Torres:2002}, and selecting the combination giving the highest
cross-correlation value averaged over all observations. Solar
metallicity was assumed, along with surface gravity ($\log g$) values
of 4.5 for both stars, which are close to our final results in
Section~\ref{sec:dimensions}.  In this way we determined temperatures
of $5930 \pm 100$~K and $5880 \pm 100$~K for the primary (the slightly
more massive star eclipsed at the deeper minimum) and secondary,
corresponding approximately to spectral types of G0 and G1, and $v
\sin i$ values of $8.4 \pm 0.8$~\kms\ and $8.2 \pm 0.5$~\kms,
respectively. Uncertainties are based on the scatter from the
individual spectra, conservatively increased to account for possible
systematic errors.  Minor differences between the adopted composition
and surface gravity compared to the final results have only a minor
effect on these values that is well within our quoted
uncertainties. Radial velocities were then determined using the
templates in our grid with parameters nearest to these values ($T_{\rm
  eff} = 6000$~K and $v \sin i = 8$~\kms\ for both stars). We also
measured the light ratio as $\ell_2/\ell_1 = 0.944 \pm 0.020$ at the
mean wavelength of our observations, 5187~\AA.

\begin{figure}
\epsscale{1.15}
\plotone{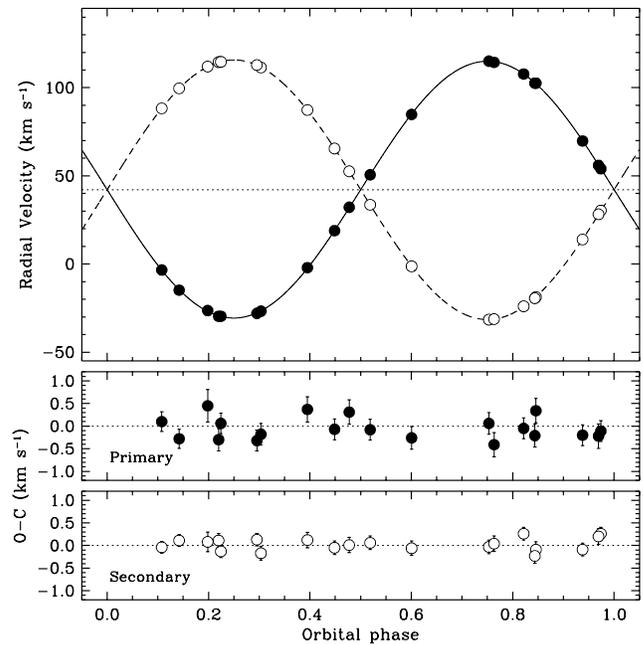}
\figcaption{Radial velocity measurements for \epic\ with our adopted
  model. Primary and secondary measurements are represented with
  filled and open circles, respectively. The dotted line marks the
  center-of-mass velocity of the system. Error bars are too small to
  be visible. They are seen in the lower panels, which display the
  residuals. Phases are counted from the reference time of primary
  eclipse.\label{fig:rvs}}
\end{figure}

The resulting heliocentric radial velocities on the native CfA
zero-point system \citep{Stefanik:1999, Stefanik:2006} are listed in
Table~\ref{tab:rvs} along with their uncertainties. The velocity
uncertainties for the primary are almost twice as large as those of
the secondary, even though the line broadening is very nearly the
same.  A visual representation of our measurements can be seen in
Figure~\ref{fig:rvs}, along with our final model described in the next
section.

\setlength{\tabcolsep}{3pt}  
\begin{deluxetable}{lccccc}
\tablewidth{0pc}
\tablecaption{Heliocentric Radial Velocity Measurements of \epic \label{tab:rvs}}
\tablehead{
\colhead{HJD} &
\colhead{$RV_1$} &
\colhead{$\sigma_1$} &
\colhead{$RV_2$} &
\colhead{$\sigma_2$} &
\colhead{Orbital}
\\
\colhead{(2,400,000$+$)} &
\colhead{(\kms)} &
\colhead{(\kms)} &
\colhead{(\kms)} &
\colhead{(\kms)} &
\colhead{phase}
}
\startdata
  57523.9695  & \phn\phn$-$2.06  &  0.29  &   \phs\phn87.35  &  0.16  &   0.3951 \\
  57526.9118  &      \phs102.58  &  0.28  &    \phn$-$18.68  &  0.16  &   0.8459 \\
  57529.8984  &    \phn$-$26.73  &  0.25  &      \phs111.44  &  0.14  &   0.3034 \\
  57535.8802  &    \phn$-$29.64  &  0.26  &      \phs114.55  &  0.15  &   0.2199 \\
  57539.9522  &      \phs102.57  &  0.26  &    \phn$-$19.37  &  0.15  &   0.8438 \\
  57550.8860  &   \phs\phn50.60  &  0.24  &   \phs\phn33.60  &  0.14  &   0.5189 \\
  57552.8622  &      \phs107.74  &  0.23  &    \phn$-$23.94  &  0.13  &   0.8216 \\
  57553.8560  &   \phs\phn53.98  &  0.23  &   \phs\phn30.35  &  0.13  &   0.9739 \\
  57554.9536  &    \phn$-$14.84  &  0.21  &   \phs\phn99.60  &  0.12  &   0.1421 \\
  57555.9545  &    \phn$-$27.99  &  0.24  &      \phs112.88  &  0.13  &   0.2954 \\
  57556.9542  &   \phs\phn18.92  &  0.24  &   \phs\phn65.52  &  0.14  &   0.4486 \\
  57557.9466  &   \phs\phn84.81  &  0.26  & \phn\phn$-$1.29  &  0.15  &   0.6006 \\
  57558.9417  &      \phs115.04  &  0.25  &    \phn$-$31.50  &  0.14  &   0.7531 \\
  57566.8803  &   \phs\phn55.94  &  0.28  &   \phs\phn28.20  &  0.16  &   0.9693 \\
  57854.9782  & \phn\phn$-$3.39  &  0.22  &   \phs\phn88.26  &  0.12  &   0.1078 \\
  57879.9771  &   \phs\phn69.77  &  0.23  &   \phs\phn13.94  &  0.13  &   0.9378 \\
  57907.9566  &    \phn$-$29.64  &  0.24  &      \phs114.67  &  0.13  &   0.2244 \\
  57920.8409  &    \phn$-$26.40  &  0.37  &      \phs112.00  &  0.21  &   0.1984 \\
  58033.6246  &   \phs\phn32.18  &  0.28  &   \phs\phn52.57  &  0.16  &   0.4776 \\
  58061.5978  &      \phs114.34  &  0.28  &    \phn$-$31.20  &  0.16  &   0.7632
\enddata
\tablecomments{Orbital phases are counted from the reference time of
  primary eclipse. Final velocity uncertainties result from scaling
  the values listed for the primary and secondary by the near-unity
  factors $f_1$ and $f_2$, respectively, from our global analysis
  described in Section~\ref{sec:lightcurve}.}
\end{deluxetable}
\setlength{\tabcolsep}{6pt}  

\section{Light curve analysis}
\label{sec:lightcurve}

Our analysis of the \ktwo\ photometry of \epic\ was based on the
Nelson-Davis-Etzel binary model \citep{Popper:1981, Etzel:1981} that
underlies the popular EBOP code and its descendants. This relatively
simple model treats the star shapes as biaxial spheroids for the
phase-dependent component luminosities, and includes a simple
reflection prescription following \cite{Binnendijk:1960}. It is
suitable for well-detached eclipsing systems such as ours in which the
stars have relatively small oblateness.  The version we use ({\tt EB})
is a rewrite by \cite{Irwin:2011}\footnote{\url
  https://github.com/mdwarfgeek/eb~.}, and is convenient for use
within the Markov Chain Monte Carlo (MCMC) scheme we apply below.  The
main adjustable variables are the orbital period ($P$) and reference
epoch of primary eclipse ($T_0$, which is strictly the time of
conjunction), the central surface brightness ratio in the
\kepler\ passband ($J \equiv J_2/J_1$), the sum of the relative radii
($r_1+r_2$) and their ratio ($k \equiv r_2/r_1$), the cosine of the
inclination angle ($\cos i$), and the eccentricity parameters $e
\cos\omega$ and $e \sin\omega$, with $e$ being the eccentricity and
$\omega$ the longitude of periastron. Limb-darkening coefficients for
the quadratic law were taken from \cite{Claret:2011} for the measured
stellar properties ($T_{\rm eff}$, $\log g$, and solar metallicity),
and were assumed to be the same for the two stars given their nearly
identical parameters. The linear and quadratic coefficients used are
$u_1 = 0.382$ and $u_2 = 0.258$.

Because our detrending procedure for the \ktwo\ photometry that is
intended to eliminate the obvious modulation due to spots also
effectively removes any other out-of-eclipse variability, effects such
as tidal distortions (ellipsoidal variability) and reflection are no
longer present in the detrended light curve. Consequently, for the
modeling we considered the stars to be spherical (setting the mass
ratio to zero, which has no effect on any other parameter) and
switched the reflection effect off in {\tt EB}.  Gravity darkening
then becomes irrelevant. As the out-of-eclipse phases yield no useful
information, we restricted the analysis to data within 0.03 in phase
from the center of each eclipse (approximately 2.5 times the total
duration of the eclipses). Additionally, to account for the finite
time of integration of the \ktwo\ long-cadence observations, the model
light curve was oversampled and then integrated over the 29.4~min
duration of each cadence before being compared with the observations
\citep[see][]{Gilliland:2010, Kipping:2010}.

As indicated earlier, the photometric aperture used to extract the
light curve from the \ktwo\ images contains a number of other stars
that contribute flux (see Figure~\ref{fig:CFHT}). We accounted for
this extra flux by including the additional third light parameter
$L_3$ (defined such that $L_1+L_2+L_3 = 1$), and we assumed the
contaminating stars have constant brightness as we only have a
measurement of their flux at a single epoch.

Our method of analysis used the {\tt emcee\/}\footnote{\url
  http://dan.iel.fm/emcee~.} code of \cite{Foreman-Mackey:2013}, which
is a Python implementation of the affine-invariant MCMC ensemble
sampler proposed by \cite{Goodman:2010}. We used 100 walkers with
chain lengths of 20,000 each, discarding the first 5,000 as burn-in.
Uniform (non-informative) or log-uniform priors over suitable ranges
were adopted for all parameters (see below), and convergence of the
chains was checked visually, requiring also a Gelman-Rubin statistic
of 1.01 or smaller for each parameter \citep{Gelman:1992}.

Despite the high quality of the \ktwo\ photometry, initial tests
revealed that the radius ratio $k$ was poorly constrained and suffered
from strong degeneracies with other parameters. This is a common
occurrence in eclipsing binaries such as \epic\ with similar
components displaying shallow, partial eclipses. In this case the two
eclipses are only about 6\% deep. The most effective remedy is to take
advantage of the fact that the radius ratio and the light ratio are
highly correlated with each other ($\ell_2/\ell_1 \propto k^2$), and
to constrain $k$ indirectly by using our spectroscopic determination
of $\ell_2/\ell_1$ from Section~\ref{sec:spectroscopy}. We did this by
applying a Gaussian prior to the light ratio. We transformed our
measured spectroscopic light ratio at 5187~\AA\ to the \kepler\ band
by applying a small correction based on the temperature difference
inferred from $J$ (see Section~\ref{sec:dimensions}) and synthetic
spectra, obtaining $\ell_2/\ell_1 = 0.96 \pm 0.02$ in $K\!p$.

We also found $L_3$ to be poorly constrained. An external estimate of
third light contamination was obtained from the brightness
measurements in Table~\ref{tab:CFHT} of all companions within the
photometric aperture (15\farcs4 in radius, see Figure~\ref{fig:CFHT}),
using the magnitudes in the Sloan $r'$ band, which is near the
\ktwo\ passband. To account for possible errors from the small
wavelength difference we conservatively inflated the companion
magnitude uncertainties so that they are individually no smaller than
0.2~mag.  We obtained $L_3 = 0.084 \pm 0.017$, and used this as a
Gaussian prior in our MCMC analysis. Because this value is dominated
by the brighter companions inside of 13\arcsec, and all the ones near
the edge of the circular aperture are very faint, the result is
completely insensitive to the treatment of partial pixels for the
latter stars.

While our spectroscopic observations gave no indication that the orbit
is anything other than circular, a preliminary light curve analysis
uncovered a very small but statistically significant displacement of
the secondary eclipse from phase 0.5, suggestive of a detectable
eccentricity. The $e \cos\omega$ parameter that measures this
displacement appeared fairly well constrained, but the complementary
parameter $e \sin\omega$ was not.  Conversely, spectroscopic
observations are typically more sensitive to $e \sin\omega$ than to $e
\cos\omega$ in systems with appreciable eccentricity, so it is
sometimes beneficial to combine the two types of measurements. In our
case the apparent eccentricity is so small that spectroscopy provides
almost no information on $e \sin\omega$. Nevertheless, we took this
approach of combining the data for our final analysis and solved for
three additional parameters: the center-of-mass velocity of the system
($\gamma$), and the velocity semi-amplitudes $K_1$ and $K_2$. Light
travel time across the binary was taken into account for completeness,
as in some systems this can also cause a displacement of the secondary
eclipse from phase 0.5, although in this case its effect is completely
negligible (a delay of less than 0.5 sec, two orders of magnitude
smaller than the measured displacement of 91 sec). The relative
weighting between the photometry and the primary and secondary radial
velocities was handled by including additional adjustable parameters
($f_{\ktwo}$, $f_1$, and $f_2$, respectively) to inflate the
observational errors. These scale factors were solved for
self-consistently and simultaneously with the other orbital quantities
\citep[see][]{Gregory:2005}. The initial error assumed for the
photometric measurements is 200~ppm, and the initial errors for the
velocities are listed in Table~\ref{tab:rvs}.  While somewhat
improved, this combined analysis still did not yield a very good
determination of $e \sin\omega$, which remained weakly
constrained. However, this does not affect the quality of the solution
because $e \sin\omega$ is essentially uncorrelated with the rest of
the geometric properties. The significance of the resulting $e
\cos\omega$ term is at the 2.8$\sigma$ level.  Attempts to solve for
the limb-darkening coefficients also did not succeed (unsurprising
given the shallow and partial nature of the eclipses), so we held them
fixed at their theoretical values.

\setlength{\tabcolsep}{4pt}
\begin{deluxetable}{lcc}
\tablewidth{0pc}
\tablecaption{Results from our Combined MCMC Analysis for \epic \label{tab:LCfit}}
\tablehead{ \colhead{~~~~~~~~~Parameter~~~~~~~~~} & \colhead{Value} & \colhead{Prior} }
\startdata
 $P$ (days)\dotfill               &  $6.527139^{+0.000013}_{-0.000013}$  & [6, 7] \\ [1ex]
 $T_0$ (HJD$-$2,400,000)\dotfill  &  $57345.15844^{+0.00016}_{-0.00016}$ & [57342, 57347] \\ [1ex]
 $J$\dotfill                      &  $0.9843^{+0.0067}_{-0.0062}$        & [0.5, 1.5] \\ [1ex]
 $r_1+r_2$\dotfill                &  $0.1107^{+0.0011}_{-0.0011}$        & [0.05, 0.20] \\ [1ex]
 $k$\dotfill                      &  $0.9878^{+0.0097}_{-0.0105}$        & [0.5, 1.5] \\ [1ex]
 $\cos i$\dotfill                 &  $0.08262^{+0.00075}_{-0.00088}$     & [0, 1] \\ [1ex]
 $e \cos\omega$\dotfill           &  $+0.000122^{+0.000044}_{-0.000044}$ & [$-$1, 1] \\ [1ex]
 $e \sin\omega$\dotfill           &  $+0.00084^{+0.00075}_{-0.00088}$    & [$-$1, 1] \\ [1ex]
 $L_3$\dotfill                    &  $0.074^{+0.020}_{-0.018}$           & [0.0, 0.2]\\ [1ex]
 $\gamma$ (\kms)\dotfill          &  $42.143^{+0.027}_{-0.034}$          & [30, 50] \\ [1ex]
 $K_1$ (\kms)\dotfill             &  $72.801^{+0.080}_{-0.073}$          & [60, 80] \\ [1ex]
 $K_2$ (\kms)\dotfill             &  $73.607^{+0.052}_{-0.047}$          & [60, 80] \\ [1ex]
 $f_{\ktwo}$\dotfill              &  $1.476^{+0.061}_{-0.065}$           & [10$^{-2}$, 10$^2$] \\ [1ex]
 $f_1$\dotfill                    &  $0.91^{+0.24}_{-0.10}$              & [10$^{-2}$, 10$^2$] \\ [1ex]
 $f_2$\dotfill                    &  $1.02^{+0.26}_{-0.12}$              & [10$^{-2}$, 10$^2$] \\ [1ex]
\multicolumn{3}{c}{Derived quantities} \\ [1ex]
 $r_1$\dotfill                    &  $0.05567^{+0.00061}_{-0.00059}$ & \nodata \\ [1ex]
 $r_2$\dotfill                    &  $0.05499^{+0.00061}_{-0.00062}$ & \nodata \\ [1ex]
 $i$ (deg)\dotfill                &  $85.261^{+0.051}_{-0.043}$ & \nodata \\ [1ex]
 $\ell_2/\ell_1$\dotfill          &  $0.958^{+0.022}_{-0.019}$ & \nodata 
\enddata
\tablecomments{The values listed correspond to the mode of the
  respective posterior distributions, and the uncertainties represent
  the 68.3\% credible intervals that include a contribution from extra
  photometric noise caused by stellar activity (see text). See also
  footnote~\ref{foot:mode}. All priors are uniform over the specified
  ranges, except those for $f_{\ktwo}$, $f_1$, and $f_2$, which are log-uniform.}
\end{deluxetable}
\setlength{\tabcolsep}{6pt}

\begin{figure*}
\epsscale{1.18}
\plotone{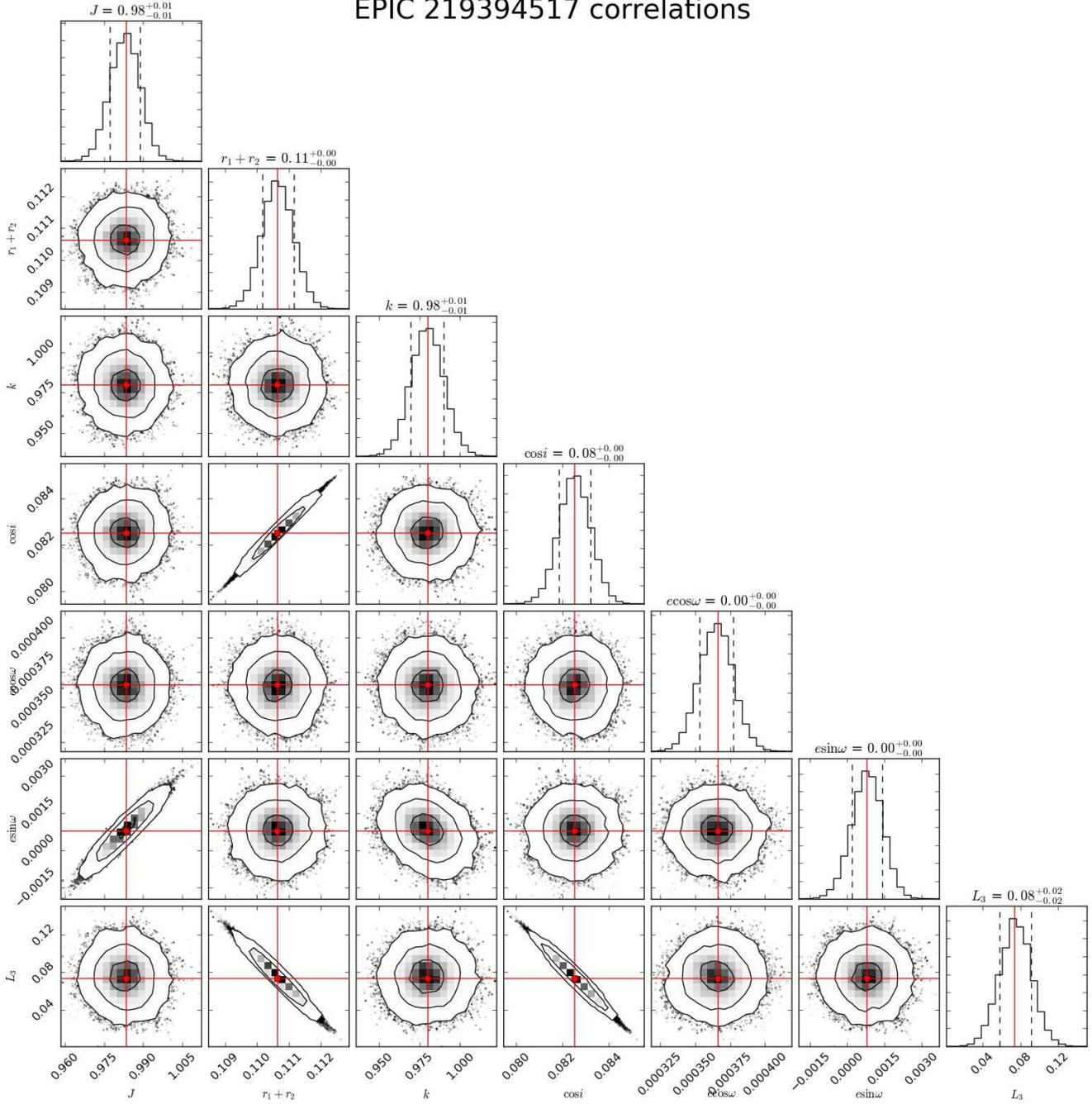}
\figcaption{``Corner plot'' \citep{Foreman-Mackey:2016}\footnote{\url
    https://github.com/dfm/corner.py~.} from the MCMC analysis of \epic\
  illustrating the correlations among a selection of the fitted
  parameters of our solution.  Contour levels correspond to 1, 2, and
  3$\sigma$, and the histograms on the diagonal represent the
  posterior distribution for each parameter, with the mode and
  internal 68.3\% confidence levels indicated. More realistic errors
  are discussed in the text.\label{fig:corner}}
\end{figure*}

The results of our analysis are given in Table~\ref{tab:LCfit}, where
we report for each parameter the mode of the corresponding posterior
distribution.  Posterior distributions for all derived quantities were
computed directly from the MCMC chains of the fitted parameters
involved.\footnote{We note that our choice to report the mode of the
  distributions will cause small but unavoidable differences between
  the reported values for some of the derived quantities and those
  that one would compute directly using the modal values of the
  originally adjusted parameters in the table.\label{foot:mode}}
Because $e \sin\omega$ is so poorly constrained we do not report a
value for either the eccentricity or $\omega$. Nevertheless, based on
its posterior distribution we can constrain the eccentricity to be
between 0.00012 and 0.0023 at the 99\% confidence level.  A few
correlations remain among the variables, particularly between
$r_1+r_2$, $L_3$, and $\cos i$, and also between $e \sin\omega$ and
$J$, as can be seen graphically in Figure~\ref{fig:corner}. The
eclipses are grazing.

\begin{figure}
\centering
\begin{tabular}{c}
\includegraphics[width=8.0cm]{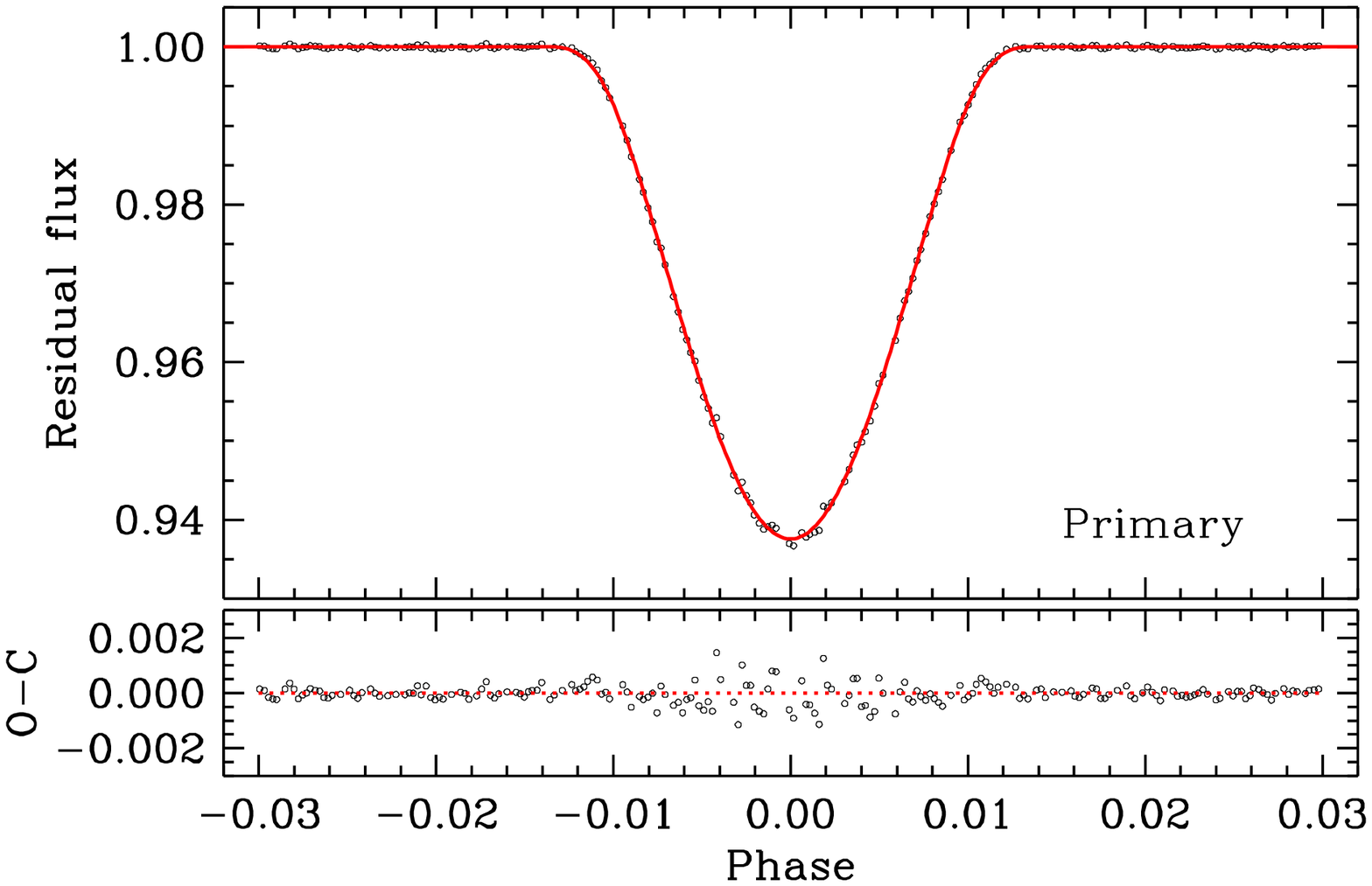} \\ [0.5ex]
\includegraphics[width=8.0cm]{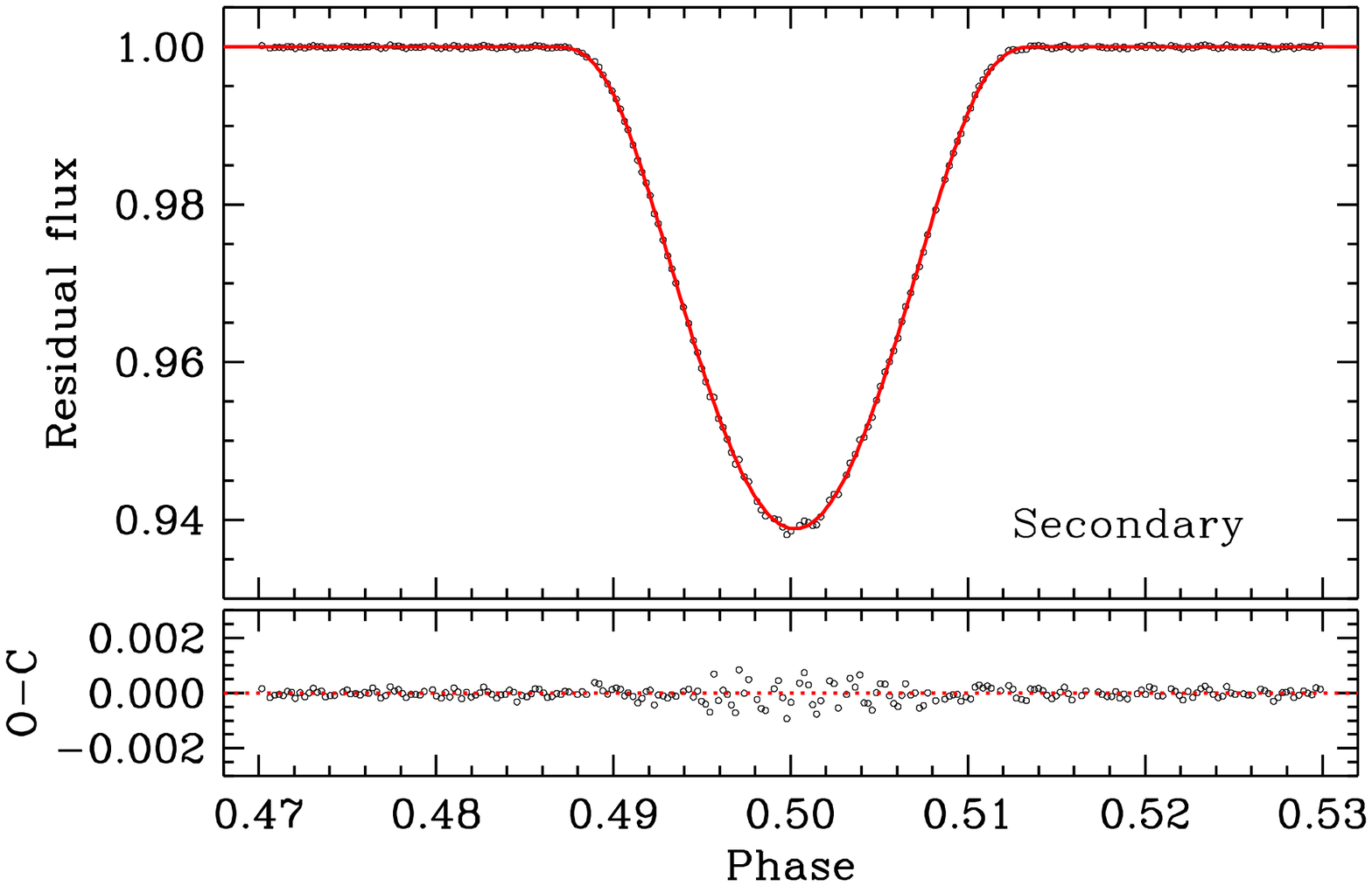}
\end{tabular}

\figcaption{\ktwo\ observations of \epic\ and our adopted model.
  Residuals are shown at the bottom for each eclipse. Note that this
  model, computed from the modal values of the parameter posterior
  distributions, is not necessarily the best-fit model in terms of the
  statistical likelihood. In practice, however, the difference is
  imperceptible to the eye.\label{fig:LCfit}}
\end{figure}

Our adopted model based on the parameters listed above is shown
graphically in Figure~\ref{fig:LCfit}, along with the
\ktwo\ observations. Both eclipses display an obvious increase in the
scatter of the residuals compared to the out-of-eclipse sections. It
is conceivable that this could be caused by the detrending procedure
we have applied (see Section~\ref{sec:photometry}), in which we
simultaneously fitted the long-term stellar/instrumental variability
with a basis spline, the \kepler\ roll systematics with cubic splines,
and the eclipses with a binary model. Subtle biases in the resulting
long-term variability basis spline correction (from possible crosstalk
with the binary model), which we use to remove those effects from the
light curve and flatten it, could in principle introduce changes in
the shape of the eclipses that would result in excess scatter at those
phases. However, tests in which we repeated the detrending masking out
the eclipses to prevent this from happening showed that the large
residuals in eclipse remain, indicating they are real, and we
attribute them to spots. A similar phenomenon is commonly seen in
active eclipsing systems.  Some degree of activity in \epic\ is in
fact expected given that the stars are rotating roughly three times
more rapidly than field G stars of this age
(Section~\ref{sec:rotation}).  Furthermore, the residual scatter at
the primary minimum is larger than at the other minimum, suggesting
that the primary is the more active star of the two, possibly due to a
difference in activity cycles given that the stars are otherwise so
similar. This may also explain its larger radial-velocity
uncertainties (almost a factor of two larger than the secondary; see
Table~\ref{tab:rvs}), and the somewhat increased error in its $v \sin
i$ value as well. The impact of this extra noise during eclipse is
that the uncertainties in our MCMC analysis will be underestimated.

To account for this we carried out a residual permutation exercise in
which we shifted the residuals from our adopted model by an arbitrary
number of time indices, added them back into the model curve at each
time of observation (with wrap-around), and performed the MCMC
analysis again on the synthetic data set. Because the residuals in
eclipse are larger than out of eclipse, the residual permutation was
done separately in these two regions to avoid producing simulated
observations with smaller in-eclipse errors on average than the actual
data, which would lead to underestimated parameter errors
\citep[see][]{Hartman:2018}. For each new MCMC analysis we
simultaneously perturbed the theoretical linear and quadratic limb
darkening coefficients by adding Gaussian noise with a standard
deviation of 0.10, and used the same perturbed coefficients for both
stars given that their temperatures are so similar. We repeated this
100 times, and adopted the scatter (standard deviation) of the
resulting distribution for each parameter as a more realistic measure
of the uncertainty that accounts for the influence of spots.  Finally,
we added these uncertainties in quadrature to the internal errors from
our original MCMC analysis to arrive at the final uncertainties
reported in Table~\ref{tab:LCfit}. The parameters in the table for
which the simulation errors dominate over the internal errors are $P$,
$T_0$, $r_1+r_2$, $e \cos\omega$, $r_1$, and $r_2$.

\section{Absolute dimensions}
\label{sec:dimensions}

Posterior distributions of the physical characteristics of the
\epic\ stars, and other system properties, were derived by directly
combining the Markov chains of the parameters on which those
properties depend. We report the mode of these distributions and their
68.3\% confidence intervals in Table~\ref{tab:dimensions}. For
properties that use information external to our MCMC analysis the
external quantities (effective temperatures, bolometric corrections,
reddening, and apparent visual magnitude) were assumed to be
distributed normally and independently for combining them with the
chains.

The chemical composition (iron abundance) of \epic\ has not been
measured directly, but we may reasonably assume it is the same as that
of the parent cluster Ruprecht~147, which is reported to be ${\rm
  [Fe/H]} = +0.10 \pm 0.04$ by \cite{Curtis:2013} on the basis of
independent but consistent spectroscopic determinations for several
dwarf and giant stars in the cluster \citep[see also Section~5.2
  of][]{Curtis:2016}.\footnote{For most of these stars the effect of
  microscopic diffusion on the surface abundance is very small ($\leq
  0.02$~dex), and we have not attempted to apply any corrections
  \citep[see][]{Dotter:2017}.} There is no evidence in those
measurements for $\alpha$-element enhancement, so we assume none.

Our adopted effective temperatures for the components are those from
our spectroscopic determinations. While the formal difference between
the primary and secondary temperatures of $50 \pm 140$~K has a large
uncertainty, $\Delta T_{\rm eff}$ is much better determined from the
difference in the eclipse depths through the $J$ parameter in our
light curve solution, which gives $25 \pm 10$~K. These two estimates
of $\Delta T_{\rm eff}$ are similar enough, and the uncertainty in the
absolute $T_{\rm eff}$ determinations large enough, that we have not
adjusted the individual values and retain the original spectroscopic
temperatures for the rest of our analysis.

Additional estimates of the mean temperature of the system may be
derived from color indices based on brightness measurements available
in a variety of passbands. The results, however, are strongly
influenced by reddening. We have chosen here to do the reverse, i.e.,
to make use of the photometric temperature estimates to infer the
reddening, by seeking agreement between those reddening-dependent
values and our spectroscopically derived mean temperature. We used
brightness measurements in the Johnson, 2MASS, and Sloan systems
\citep{Henden:2014, Henden:2015, Skrutskie:2006, Curtis:2013} to
construct nine non-independent color indices.  Color-temperature
calibrations from \cite{Casagrande:2010} and \cite{Huang:2015} were
then used to derive a photometric temperature from each index and each
available calibration for that index, resulting in a total of thirteen
$T_{\rm eff}$ values.  Both sets of color-temperature calibrations
include terms that depend on metallicity, for which we adopted the
above value of ${\rm [Fe/H]} = +0.10 \pm 0.04$.  To each index we
applied a wavelength-dependent reddening correction as prescribed by
\cite{Cardelli:1989}, and varied the amount of reddening until we
found agreement between the average of the 13 photometric $T_{\rm
  eff}$ estimates and the luminosity-weighted spectroscopic mean
temperature of the system. In this way we arrived at a reddening
estimate along the sightline to \epic\ of $E(B-V) = 0.112 \pm
0.029$~mag.  The corresponding visual extinction assuming $R_V = 3.1$
is $A_V = 0.347 \pm 0.090$~mag, marginally larger than the value of
$A_V = 0.25 \pm 0.05$~mag favored by \cite{Curtis:2013} and other
estimates therein.

We point out that temperatures derived from the \cite{Casagrande:2010}
and \cite{Huang:2015} calibrations differ in a systematic way, and
this has an impact on the inferred reddening (and distance). There is
considerable debate about the absolute effective temperature scale,
which ultimately relies on interferometric angular diameter
measurements, some of which have been suspected of being afflicted by
systematic errors \cite[see the above references for opposing views,
  as well as][and references therein]{Huber:2017}. Zero point
differences in the $T_{\rm eff}$ scale among various authors remain at
the $\sim$100~K level. While the calibrations of \cite{Huang:2015} are
based on angular diameters, those of \cite{Casagrande:2010} employ the
Infrared Flux Method (IRFM), and yield temperatures systematically
hotter by about 130~K.  Here we have used the temperature scale of
\cite{Huang:2015}, largely on the basis of the better agreement with
other information presented in Section~\ref{sec:models}. Accordingly,
the temperatures from \cite{Casagrande:2010} have been adjusted by
$-130$~K prior to averaging them with those of \cite{Huang:2015}. We
note, however, that there is also some recent evidence
\citep{Huber:2017, White:2018} suggesting that the IRFM of
\cite{Casagrande:2010} may be somewhat more consistent with
asteroseismic results as well as with spectroscopic determinations
based on the excitation balance of iron lines, so the debate is not
yet settled. We return to our reddening estimate in
Section~\ref{sec:models}.

As an independent check on the differential extinction we measured the
equivalent width of the interstellar \ion{Na}{1}~D1 line in six of our
spectra in which the feature is well resolved from the stellar
components. The strength of this line has been found to correlate with
the amount of extinction \citep[see, e.g.,][]{Munari:1997}, albeit
with considerable scatter. Based on a mean equivalent width of $0.27
\pm 0.02$~\AA\ and the calibration from the above authors we infer a
value of $E(B-V) \approx 0.10$~mag, consistent with our result above.

With our adopted reddening estimate the distance we derive for
\epic\ is $283^{+18}_{-16}$~pc, based on the apparent out-of-eclipse
visual magnitude of the system \citep[$V = 11.444 \pm
  0.034$;][]{Henden:2014, Henden:2015} and bolometric corrections from
\cite{Flower:1996}.  The corresponding parallax, $\pi =
3.51^{+0.23}_{-0.19}$~mas, is consistent with the {\it Gaia}/DR2
estimate of $\pi_{Gaia} = 3.310 \pm 0.040$~mas \citep{Gaia:2016,
  Gaia:2018} within about 1$\sigma$.  For completeness we note that
adopting the temperature scale of \cite{Casagrande:2010} instead of
that of \cite{Huang:2015} leads to a smaller reddening of $E(B-V) =
0.076 \pm 0.029$~mag ($A_V = 0.236 \pm 0.090$~mag), and a somewhat
reduced parallax of $\pi = 3.34^{+0.22}_{-0.19}$~mas, or a distance of
$297^{+20}_{-16}$~pc.

\begin{deluxetable}{lcc}
\tablewidth{0pc}
\tablecaption{Physical Properties of \epic \label{tab:dimensions}}
\tablehead{ \colhead{~~~~~~~~~~Parameter~~~~~~~~~~} & \colhead{Primary} & \colhead{Secondary} }
\startdata
 $M$ ($\mathcal{M}_{\sun}^{\rm N}$)\dotfill   &  $1.0782^{+0.0019}_{-0.0019}$  &  $1.0661^{+0.0027}_{-0.0021}$ \\ [1ex]
 $R$ ($\mathcal{R}_{\sun}^{\rm N}$)\dotfill   &  $1.055^{+0.011}_{-0.011}$  &  $1.042^{+0.012}_{-0.012}$ \\ [1ex]
 $q \equiv M_2/M_1$\dotfill               &          \multicolumn{2}{c}{$0.9890^{+0.0013}_{-0.0012}$} \\ [1ex]
 $a$ ($\mathcal{R}_{\sun}^{\rm N}$)\dotfill                 &          \multicolumn{2}{c}{$18.954^{+0.012}_{-0.012}$} \\ [1ex]
 $\log g$ (dex)\dotfill                   &  $4.4247^{+0.0091}_{-0.0097}$  &  $4.4303^{+0.0098}_{-0.0095}$ \\ [1ex]
 $T_{\rm eff}$ (K)\dotfill                &  5930~$\pm$~100                &  5880~$\pm$~100 \\ [1ex]
 $L$ ($L_{\sun}$)\dotfill                 &  $1.233^{+0.098}_{-0.078}$     &  $1.165^{+0.090}_{-0.078}$ \\ [1ex]
 $M_{\rm bol}$ (mag)\dotfill              &  $4.495^{+0.080}_{-0.075}$     &  $4.560^{+0.081}_{-0.075}$ \\ [1ex]
 $BC_V$ (mag)\dotfill                     &  $-0.055 \pm 0.100$            &  $-0.063 \pm 0.100$ \\ [1ex]
 $M_V$ (mag)\dotfill                      &  $4.55^{+0.13}_{-0.12}$        &  $4.62^{+0.13}_{-0.12}$ \\ [1ex]
 $v_{\rm sync} \sin i$ (\kms)\tablenotemark{a}\dotfill     &  $8.152^{+0.088}_{-0.088}$     &  $8.053^{+0.088}_{-0.093}$ \\ [1ex]
 $v \sin i$ (\kms)\tablenotemark{b}\dotfill                &  $8.4 \pm 0.8$                 &  $8.2 \pm 0.5$ \\ [1ex]
 $E(B-V)$ (mag)\dotfill                   &          \multicolumn{2}{c}{0.112~$\pm$~0.029} \\ [1ex]
 $A_V$ (mag)\dotfill                      &          \multicolumn{2}{c}{0.347~$\pm$~0.090} \\ [1ex]
 Dist.\ modulus (mag)\dotfill             &          \multicolumn{2}{c}{$7.26^{+0.13}_{-0.13}$} \\ [1ex]
 Distance (pc)\dotfill                    &          \multicolumn{2}{c}{$283^{+18}_{-16}$} \\ [1ex]
 $\pi$ (mas)\dotfill                      &          \multicolumn{2}{c}{$3.51^{+0.23}_{-0.19}$} \\ [1ex]
 $\pi_{Gaia/{\rm DR2}}$ (mas)\dotfill     &          \multicolumn{2}{c}{$3.310 \pm 0.040$}
\enddata
\tablecomments{The masses, radii, and semimajor axis $a$ are expressed
  in units of the nominal solar mass and radius
  ($\mathcal{M}_{\sun}^{\rm N}$, $\mathcal{R}_{\sun}^{\rm N}$) as
  recommended by 2015 IAU Resolution B3 \citep[see][]{Prsa:2016}, and
  the adopted solar temperature is 5772~K (2015 IAU Resolution
  B2). Bolometric corrections are from the work of \cite{Flower:1996},
  with conservative uncertainties of 0.1~mag, and the bolometric
  magnitude adopted for the Sun appropriate for this $BC_V$ scale is
  $M_{\rm bol}^{\sun} = 4.732$ \citep[see][]{Torres:2010b}. See text
  for the source of the reddening. For the apparent visual magnitude
  of \epic\ out of eclipse we used $V = 11.444 \pm 0.034$
  \citep{Henden:2014, Henden:2015}.  See footnote~\ref{foot:mode} for
  an explanation of slight inconsistencies that may be present in the
  values reported here, which correspond to the mode of the posterior
  distribution of each parameter.}
\tablenotetext{a}{Synchronous projected rotational velocity assuming
  spin-orbit alignment.}
\tablenotetext{b}{Measured projected rotational velocity.}
\end{deluxetable}

Table~\ref{tab:dimensions} includes the predicted projected rotational
velocities assuming spin-orbit alignment and synchronous rotation
($v_{\rm sync}\sin i$), which are seen to be consistent with the
spectroscopically measured $v \sin i$ values. Under the same
assumptions the ratio (secondary/primary) of the $v \sin i$ values,
$0.98 \pm 0.11$, is a direct measure of the radius ratio $k$, and
agrees with the much more precise value from the light curve analysis.

\section{Rotation and activity}
\label{sec:rotation}

\begin{figure}
\epsscale{1.15}
\plotone{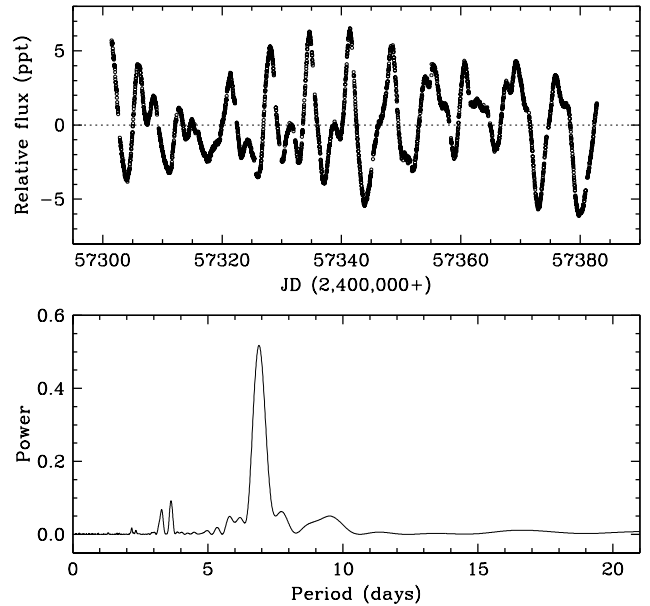}

\figcaption{\emph{Top:} Rotational modulation in the light curve of
  \epic\ (relative flux in parts per thousand, ppt). Eclipses have
  been removed for clarity, along with a long-term instrumental
  drift. \emph{Bottom:} Lomb-Scargle periodogram of the observations
  in the top panel, showing a dominant peak at $P_{\rm rot} = 6.89 \pm
  0.27$~days. \label{fig:spots}}
\end{figure}

The brightness of \epic\ varies continuously with a total amplitude of
about 1\%, for which the usual interpretation is rotational modulation
by spots on the surface of one or both stars. The variation is rather
irregular (non-sinusoidal), suggesting the spots may be moving or
changing significantly with time over the span of just a few rotation
periods. This is illustrated in Figure~\ref{fig:spots}, where we have
removed the eclipse variations as well as a long-term drift that is
most likely of instrumental nature. The lower panel shows the
Lomb-Scargle periodogram of the data with a dominant peak
corresponding to a period of $P_{\rm rot} = 6.89 \pm 0.27$~days. The
uncertainty was calculated from the half width of the peak at half
height. The same result for $P_{\rm rot}$ was obtained from an
autocorrelation analysis following \cite{McQuillan:2013}. This period
is marginally ($\sim$5\%) longer than the orbital period, which leaves
open the possibility that the rotation is not quite synchronized with
the orbital motion, as we would have expected. The estimated timescale
for this to happen for stars with convective envelopes is only
$\sim$18~Myr \citep[e.g.,][]{Hilditch:2001}, which is more than two
orders of magnitude shorter than the cluster age. Our
spectroscopically measured projected rotational velocities are not
precise enough to distinguish between synchronous and non-synchronous
rotation.  Alternatively, an apparently longer rotational period could
also be explained if there is solar-like differential rotation and the
spots occur predominantly at high latitudes.

We noted earlier that the primary appears to be the more active star
of the two, judging by the larger radial-velocity uncertainties and
the larger scatter of the light curve residuals during primary
eclipse.  As it is also slightly brighter than the secondary, it seems
more likely that the rotation period we measure corresponds to the
primary star, although variability in the secondary is probably also
present. The very similar $v \sin i$ values we measure imply similar
rotation periods, which makes it unlikely the pattern of modulation we
see is the result of the beating of two very different frequencies. We
see no obvious spectroscopic indicators of activity in our spectra,
nor does \epic\ appear to have been detected as an X-ray source
\citep[e.g., by ROSAT;][]{Voges:1999} or as an ultraviolet source
\citep[GALEX;][]{Bianchi:2011}.

\begin{figure*}
\epsscale{1.18}
\plotone{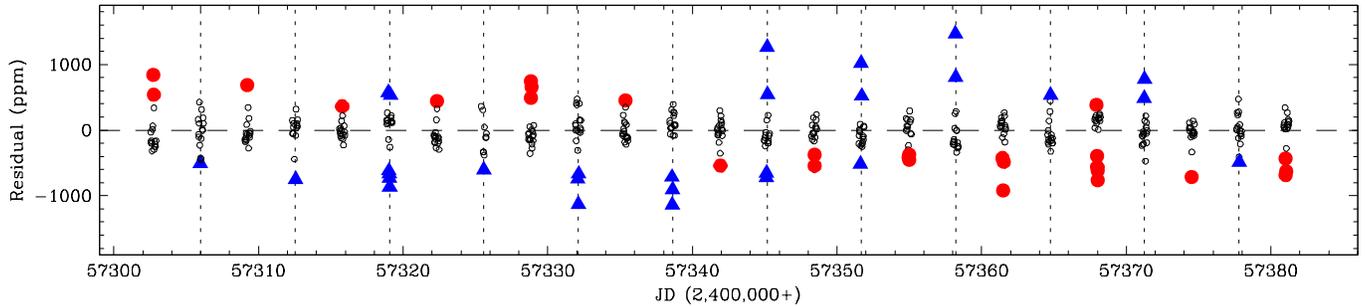}

\figcaption{Residuals of the \ktwo\ observations from our light curve
  model during primary and secondary eclipse. Primary eclipses are
  marked with dotted lines. Residuals smaller than a certain threshold
  are represented with small dots, and larger ones are shown as blue
  triangles for the primary and red circles for the secondary
  eclipses. The thresholds were set for display purposes to 360~ppm
  for the secondary and 480~ppm for the more active primary. The
  largest residual excursions for the primary are seen mostly on the
  negative side for the first half of the observations, and mostly on
  the positive side thereafter. The residuals during secondary eclipse
  show the opposite behavior.\label{fig:resid}}
\end{figure*}

Examination of the light curve residuals as a function of time has
revealed a pattern such that occasional dips in the residuals during
primary eclipse occur mostly during the first half of the
\ktwo\ observations, while during the second half the primary
residuals show outliers that are mostly positive. Exactly the opposite
is seen for the secondary, as shown in Figure~\ref{fig:resid}. If the
spots are assumed to be dark, positive residuals would correspond to
instances in which spots or spot regions on the background star are
occulted by the foreground star, resulting in a temporary reduction of
the eclipse depth because the surface brightness in the eclipsed area
is lower than would be inferred from the out-of-eclipse baseline
level. Conversely, negative residuals would result from dark spots not
covered by the foreground star. The interpretation would be reversed
if the spots were bright (faculae). This evolution in the pattern of
residuals is another sign that spots on both stars seem to be changing
on relatively short timescales of a few weeks. This is consistent with
the spot decay times measured by \cite{Giles:2017} in active stars of
this spectral type observed by \kepler, having similar levels of
photometric variability as \epic.

\section{Comparison with stellar evolution theory}
\label{sec:models}

\begin{figure}
\centering
\begin{tabular}{cc}
\includegraphics[width=4.59cm]{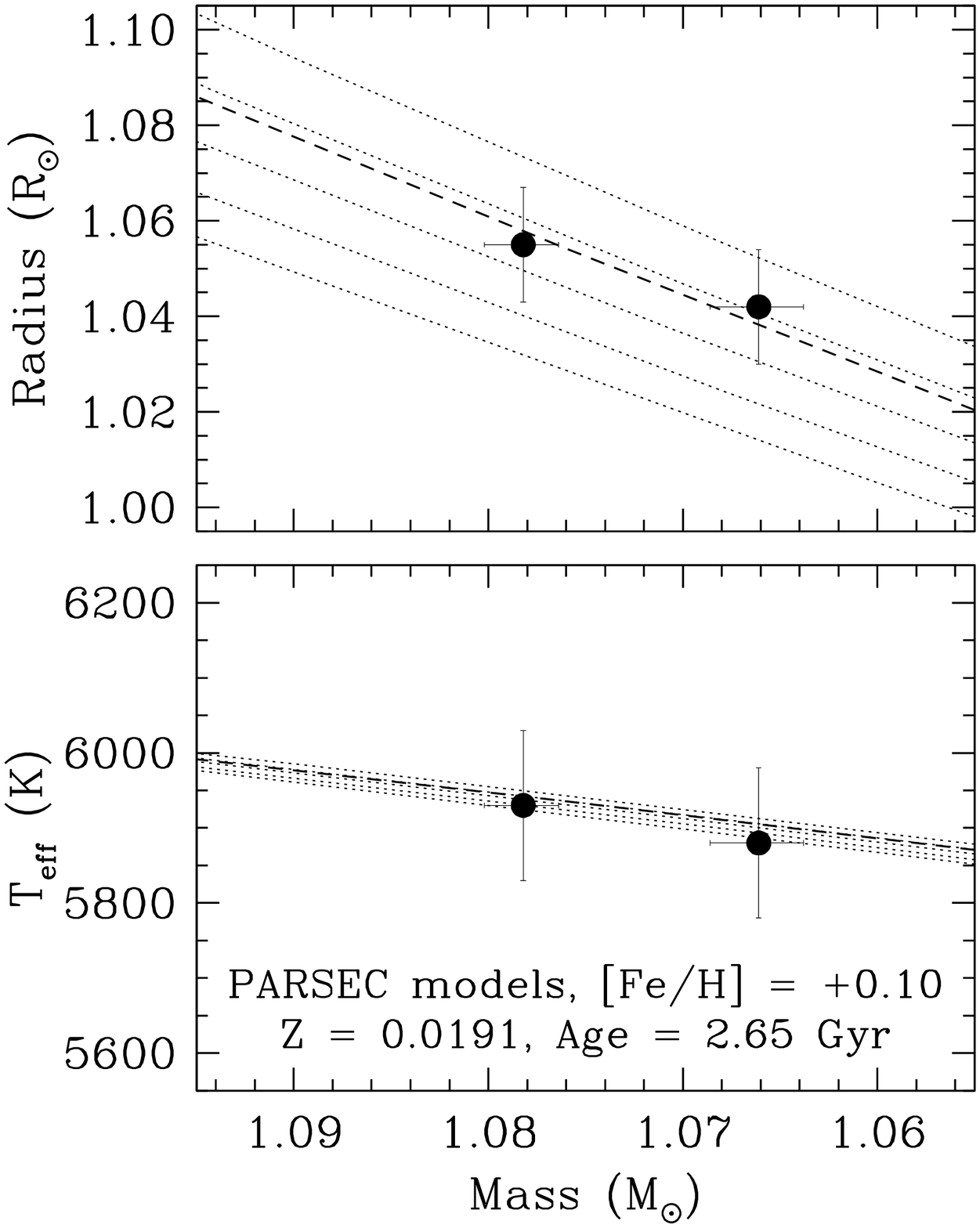} &
\includegraphics[width=3.703cm]{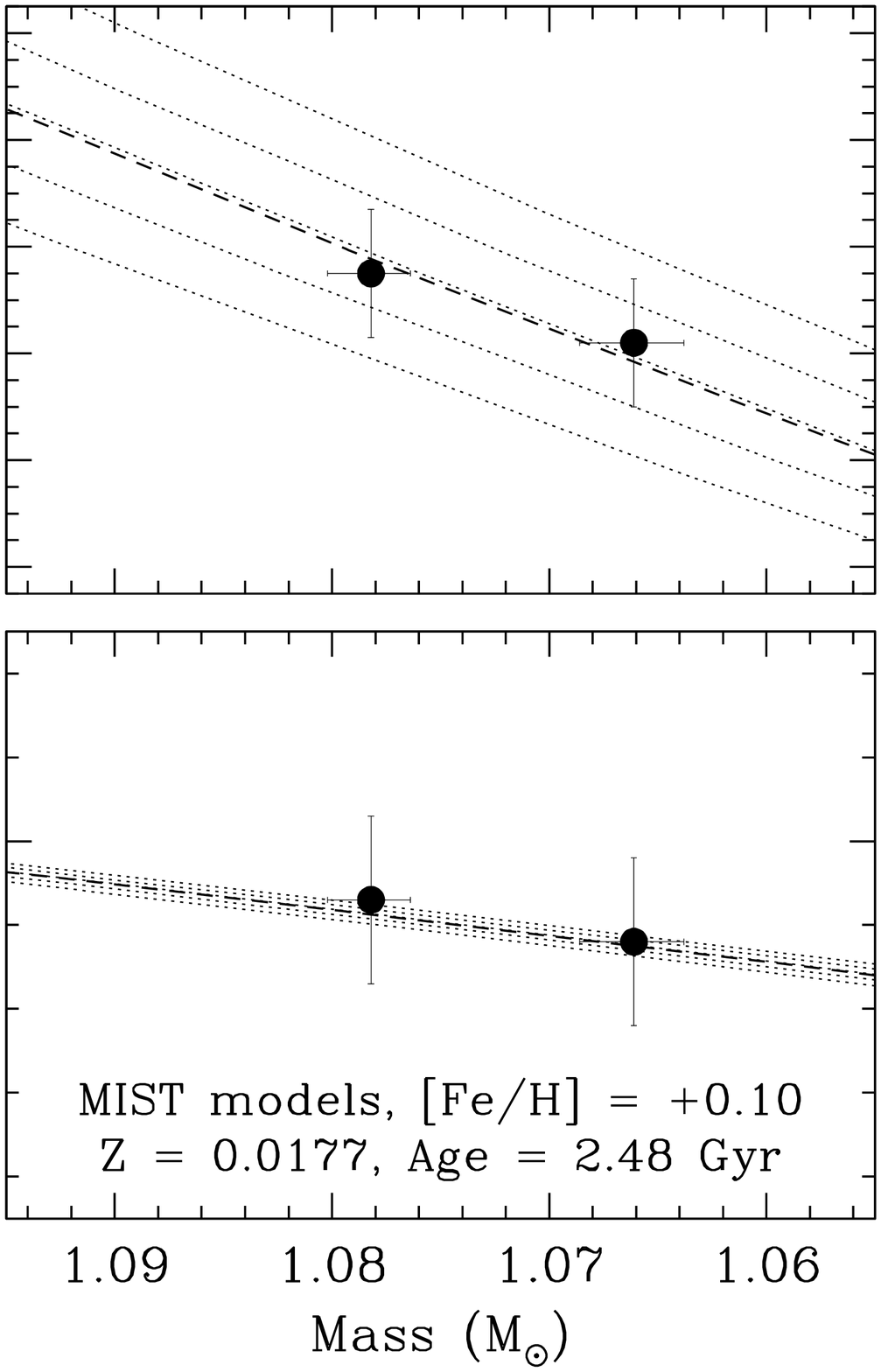}
\end{tabular}
\figcaption{Comparison of the measured masses, radii, and effective
  temperatures of \epic\ against stellar evolution models from the
  PARSEC (left) and MIST series (right). Dotted lines in all panels
  correspond to model isochrones from 2 to 3~Gyr in equally spaced
  logarithmic intervals, and the best-fit age for each model (2.65~Gyr
  for PARSEC and 2.48~Gyr for MIST) is indicated with the heavier
  dashed line. While the adopted iron abundance for both sets is the same
  (${\rm [Fe/H]} = +0.10$), the $Z$ values are not because of
  differences in the assumed solar abundance in each
  case. \label{fig:MR}}
\end{figure}

Our highly precise mass and radius determinations for \epic, with
relative errors of about 0.2\% and 1\%, respectively, offer an
opportunity for a stringent test of stellar evolution models in an
open cluster of known metallicity. In the left panels of
Figure~\ref{fig:MR} we present a comparison of our mass, radius, and
temperature determinations with isochrones based on the PAdova-TRieste
Stellar Evolution Code (version 1.2S) \citep[PARSEC;][]{Chen:2014} for
our adopted metallicity of ${\rm [Fe/H]} = +0.10$, corresponding to $Z
= 0.0191$ in these models. We find excellent agreement in both radius
and temperature at the measured masses for an age of about $2.65 \pm
0.25 \pm 0.13$~Gyr (based on the radii), where the two confidence
intervals come from the radius and metallicity errors, respectively.
Similarly good fits shown in the right panels of the figure are found
to models from the MESA Isochrones and Stellar Tracks series
\citep[MIST;][]{Choi:2016}, which is based on the Modules for
Experiments in Stellar Astrophysics package
\citep[MESA;][]{Paxton:2011, Paxton:2013, Paxton:2015}. In this case
the best-fit age for the same measured iron abundance (corresponding
to $Z = 0.0177$ in these models) is found to be $2.48 \pm 0.30 \pm
0.13$~Gyr, which is about 6\% younger. Age differences between models
are due in part to the different $Z$ values that depend on the adopted
solar abundance in each case, and to other differences in the physical
ingredients.

\setlength{\tabcolsep}{1pt}
\setlength{\tabcolsep}{6pt}

Membership of \epic\ in Ruprecht~147 allows for an interesting check
of the models via the color-magnitude diagram that is not available
for eclipsing binaries in the field. While brightness measurements in
clusters are commonly used along with model isochrones to infer the
age, distance modulus, reddening, and sometimes also the metallicity
of the population, this type of analysis is often very challenging.
Systematic errors can result from strong degeneracies usually present
among the fitted model parameters, and there are also unavoidable
complications having to do with contamination from field stars as well
as unrecognized binarity.

With the recent release of the {\it Gaia}/DR2 catalog the problem of
field star contamination has largely been eliminated from CMD studies,
or at least significantly reduced, and most importantly, the distance
modulus can now be considered to be known as high-precision parallaxes
are available for nearly every star. In Ruprecht~147 the precision of
the parallaxes is such that the depth of the cluster is resolved for
many of its members, which helps to reduce scatter in the CMD.
Reddening remains an adjustable parameter in isochrone fitting, and
binarity (especially near the turnoff) can still cause biases in the
age estimates.

\begin{figure}
\epsscale{1.15}
\plotone{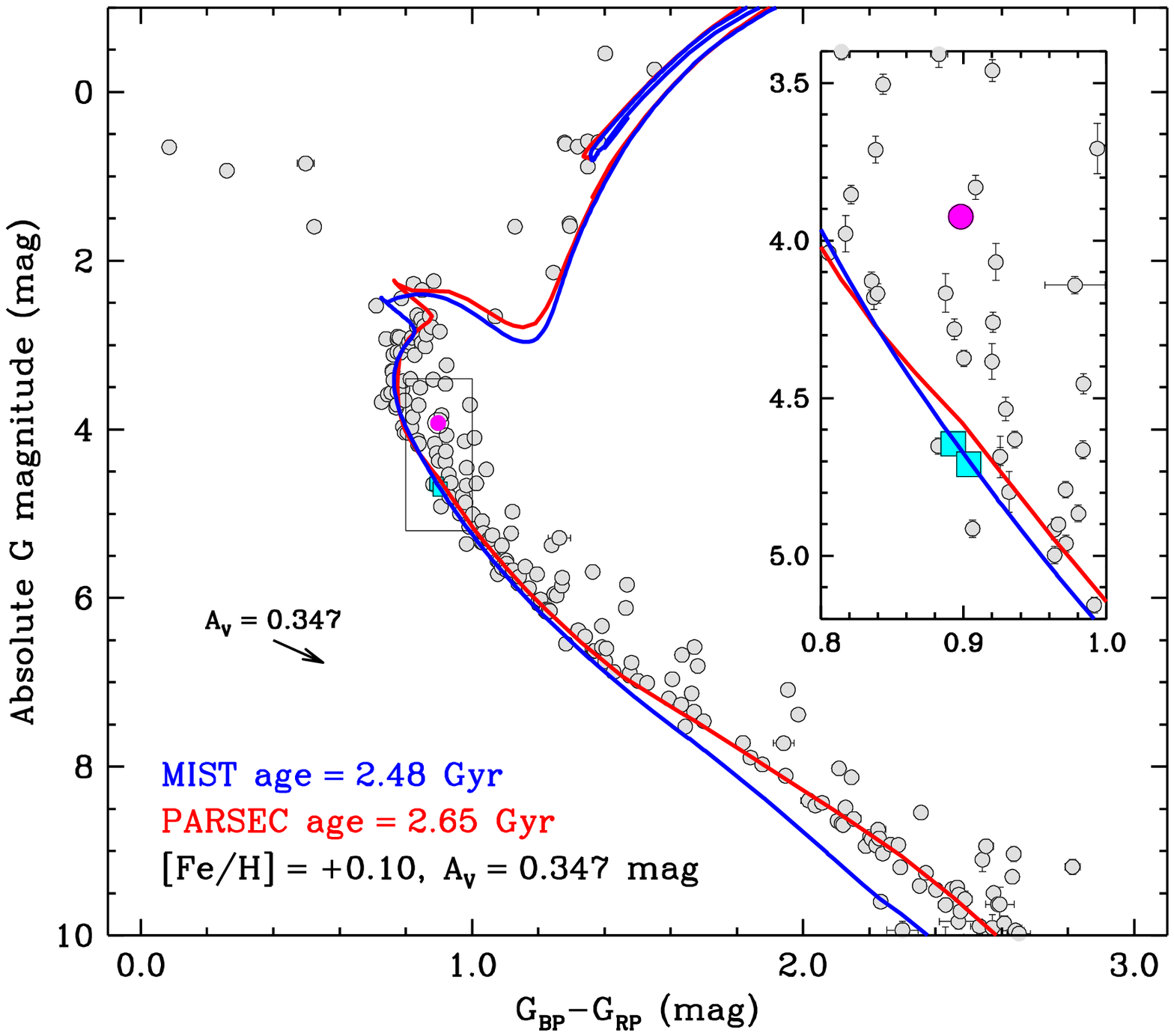}

\figcaption{Color-magnitude diagram of Ruprecht~147 members based on
  $G$ magnitudes, $G_{\rm BP}-G_{\rm RP}$ colors, and parallaxes from
  the {\it Gaia}/DR2 catalog, compared with isochrones from the MIST
  and PARSEC series \citep{Choi:2016, Chen:2014} that best match the
  binary parameters of \epic. Error bars for the measurements are
  generally too small to be visible, except in the inset. No
  parameters have been adjusted for this comparison: the isochrone
  ages are held fixed at the best-fit values derived for the binary
  (2.65~Gyr for PARSEC, 2.48~Gyr for MIST; see Figure~\ref{fig:MR}),
  the reddening of $E(B-V) = 0.112$~mag ($A_V = 0.347$~mag) applied
  directly to the isochrones is also fixed from our determination
  based on the spectroscopic temperatures and color-temperature
  calibrations (see text), and the adopted metallicity ${\rm [Fe/H]} =
  +0.10$ is the measured value for the cluster \citep{Curtis:2013}.
  The arrow represents the reddening vector.  The inset shows an
  enlargement around the position of \epic, where we have deconvolved
  the photometry (based on our measured light ratio in the $K\!p$ band
  from Table~\ref{tab:LCfit} converted to the {\it Gaia} passbands) to
  show the location of the individual components of the binary (cyan
  squares), as well as the combined light (magenta
  circle). \label{fig:CMD1}}
\end{figure}

Eclipsing binaries constrain models in a completely different way,
independent of the above complications. Our analysis of \epic\ has
provided robust (but model-dependent) age estimates, as well as a
measure of the reddening in the direction of the binary.\footnote{As
  pointed out by \cite{Curtis:2013}, extinction across Ruprecht~147
  may be patchy, which would add scatter to the
  CMD.\label{foot:patchy}} As the cluster metallicity is also known,
we are in a position to compare our best-fit isochrones from PARSEC
and MIST directly with brightness measurements of cluster members
observed by {\it Gaia} in the native $G$, $G_{\rm BP}$, and $G_{\rm
  RP}$ passbands, {\it with no free parameters}. We present such a
comparison in Figure~\ref{fig:CMD1}, based on a preliminary list of
cluster members selected according to position, proper motion,
parallax, and radial velocity criteria described in more detail
elsewhere (Curtis et al., in prep.).  The magnitudes (adjusted for
each star's distance modulus) and colors are the measured values, and
the model isochrones have been transformed to the observational plane
by applying the appropriate corrections for extinction/reddening based
on $A_V = 0.347$~mag.

The agreement in the turnoff region (most sensitive to age) and in the
giant clump is quite satisfactory for both models, while the lower
main sequence is clearly better matched by the PARSEC model. The
improved fit to these fainter stars is due to empirical corrections
described by \cite{Chen:2014} that were applied to the relation
between temperature and the Rosseland mean optical depth (the
$T$--$\tau$ relation) in order to force agreement with the masses and
radii of stars in the lower main sequence. As those authors showed,
the same corrections also improved the fits to the CMDs of several
open and globular clusters. It is therefore not surprising to find
better agreement also with Ruprecht~147. An enlargement in the area of
\epic\ with the location of the individual components is shown in the
inset. The combined light was deconvolved using our measured
$K\!p$-band light ratio from Table~\ref{tab:LCfit} transformed to the
{\it Gaia} passbands.

As noted in Section~\ref{sec:dimensions}, our estimate of the
extinction is dependent on a choice for the absolute temperature scale
implicit in the color-temperature relations we used
(\citealt{Casagrande:2010}, or \citealt{Huang:2015}). On the
assumption that extinction is uniform across the cluster, the good
agreement between models and the brightness measurements in
Figure~\ref{fig:CMD1} favors the larger extinction value of $A_V =
0.347$~mag based on the \cite{Huang:2015} zero point over the lower
value of $A_V = 0.236$~mag derived from \cite{Casagrande:2010}. The
latter choice would make the isochrones about 0.09~mag brighter and
$\sim$0.05~mag bluer, resulting in a small but visible shift compared
to the observations, most of which would be left to the right of the
models. If, however, the extinction toward \epic\ happens to be
smaller than the average for the cluster, then it is possible the
temperature scale of \cite{Casagrande:2010} is more accurate (see
footnote~\ref{foot:patchy}).

\begin{figure}
\epsscale{1.15}
\plotone{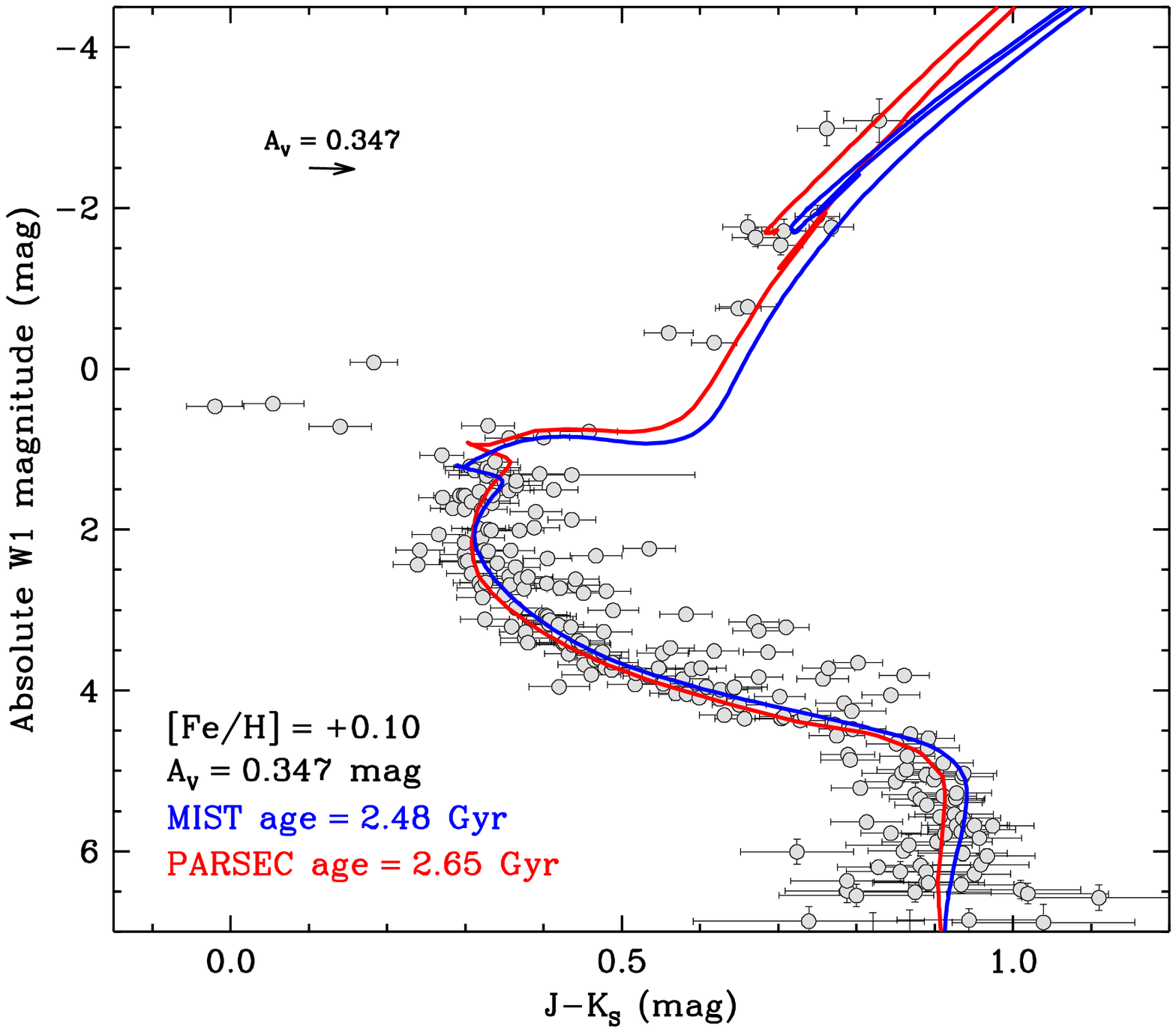}

\figcaption{Same as Figure~\ref{fig:CMD1} using near-infrared
  brightness measurements from 2MASS and WISE ($W1$ passband at
  3.4~$\mu$m). \label{fig:CMD2}}
\end{figure}

Figure~\ref{fig:CMD2} shows the CMD compared with the same model
isochrones as before in near-infrared bands, using measurements from
the WISE and 2MASS catalogs \citep{Wright:2010, Skrutskie:2006}. Once
again the agreement is very good, though somewhat better for PARSEC at
the giant branch and lower main sequence.

While an independent, detailed isochrone fitting analysis of the
cluster CMD to determine reddening and age is beyond the scope of this
work, we note, reassuringly, that preliminary results from such an
exercise with the same PARSEC and MIST models used above yield in each
case essentially the same ages we obtained here (Curtis et al., in
prep.).

\section{Discussion}
\label{sec:discussion}

With our accurate determination of the masses, radii, and effective
temperatures of its components, \epic\ joins the ranks of the
eclipsing binaries with the best measured properties
\citep[e.g.,][]{Torres:2010a}.  The stars are slightly evolved from
the zero-age main sequence, and both appear to be somewhat active,
especially the primary. Despite the relatively short period of the
system there is a hint of a very small orbital eccentricity. The
theory of equilibrium tides \citep{Zahn:1977, Zahn:1989, Claret:1997}
predicts a timescale for tidal forces to circularize the orbit of
convective stars such as these of roughly 20~Gyr
\citep{Hilditch:2001}, which would not be inconsistent with the
observation. However, other empirical evidence suggests tidal forces
are more efficient than predicted \citep[see, e.g.][]{Meibom:2005},
which would imply a shorter timescale. The comparison of our
measurements against stellar evolution models in the mass-radius
diagram yields fairly precise but model-dependent estimates of the age
of $2.65 \pm 0.25$~Gyr for PARSEC and $2.48 \pm 0.30$~Gyr for MIST,
with an additional uncertainty of 0.13~Gyr in each coming from the
metallicity error. Predictions based on the same best-fit models also
agree very well with our measured effective temperatures, which are
less fundamental than the masses and radii.  The temperatures, in
turn, when combined with available standard photometry for the system,
lead to an estimate of the extinction toward \epic\ that is similar to
previous estimates but is determined in a completely different way.

What distinguishes \epic\ from most other field eclipsing binaries
with similarly accurate properties is its membership in a star
cluster.  This enables much more stringent tests of stellar evolution
models, a better characterization of the parent population, and also
valuable cross-validation of other techniques for dating stars based,
e.g., on their oscillation frequencies or rotation periods.
\epic\ belongs to Ruprecht~147 \citep{Curtis:2013}, a relatively
nearby (300~pc) middle-aged open cluster observed by \ktwo, with a
known metallicity determined spectroscopically from about half a dozen
of its members. Using our estimate of the extinction, along with {\it
  Gaia}/DR2 parallaxes of cluster members, we find that the same model
isochrones from PARSEC and MIST that reproduce the binary measurements
in the mass-radius diagram also provide a reasonably good fit in the
color-magnitude diagram of the cluster, particularly for the PARSEC
model, both in the optical ($G$, $G_{\rm BP}$, and $G_{\rm RP}$
passbands from {\it Gaia}) and in the near infrared ($W1$, $J$, and
$K_S$ passbands from WISE and 2MASS). The models therefore appear to
be self-consistent in terms of their predictions of the internal
structure, on the one hand, and the passband-specific fluxes, on the
other, which depend on the adopted bolometric corrections,
color/temperature transformations, and model atmospheres. We emphasize
once again that the comparison in the CMD has no adjustable
parameters: the age and extinction are determined separately based on
the binary observations, the distance to all members has recently been
supplied by the {\it Gaia\/} mission, and the metallicity is a known
quantity.  Few such tests have been available in the past because
well-studied eclipsing binaries in clusters are rare.

Remarkably, at least four additional relatively bright ($K\!p < 13$)
detached eclipsing binaries that are amenable to study have been found
in Ruprecht~147 and have high-quality space-based light curves, a
situation that to our knowledge is unprecedented for an open cluster.
\epic\ is the first of these eclipsing binaries to be analyzed.
Spectroscopic observations of the others are well underway, and
complete studies of them will be the subject of future publications by
our team. Together, the components of these five binaries span a
factor of two in mass ($\sim$0.7--1.4~$M_{\sun}$) reaching up to the
turnoff region. Provided a similar precision as in \epic\ can be
achieved for the masses and radii of the other four systems, a
uniquely strong test of models in the mass-radius diagram will become
possible, including perhaps a determination of the helium abundance in
Ruprecht~147, given that the metallicity is known \citep[see,
  e.g.,][]{Brogaard:2011, Brogaard:2012}. We expect the age of the
cluster could be determined considerably more precisely than we have
here by using all five binaries simultaneously, and this could in turn
serve as an important calibrator for other methods of estimating ages.
For example, the \ktwo\ mission collected short-cadence photometric
observations suitable for asteroseismic studies of about two dozen
stars near the turnoff, as well as for a number of giants.\footnote{As
  reported in \url
  https://keplerscience.arc.nasa.gov/data/K2-programs/GO7035.txt~.}
Seismic ages from these data, when they become available, could be
compared against the binary age. Furthermore, \ktwo\ has provided a
wealth of information for measuring rotation periods in Ruprecht~147.
With a precise age determined from \epic\ and the other eclipsing
binaries, the rotational sequence (mass-$P_{\rm rot}$ relation) for
Ruprecht~147 could establish the cluster as a new reference point for
gyrochronology \citep{Barnes:2007} at an age similar to that of
NGC~6819 (2.5~Gyr), allowing for a valuable comparison between two old
clusters \citep[see][]{Meibom:2015}.

\begin{acknowledgements}

The spectroscopic observations of \epic\ were gathered with the help
of P.\ Berlind, M.\ Calkins, G.\ Esquerdo, and D.\ Latham. J.\ Mink is
thanked for maintaining the CfA echelle database. We are also grateful
to J.\ Irwin for helpful conversations about light curve solutions,
and to A.\ Dotter and L.\ Girardi for discussions about the MIST and
PARSEC stellar evolution models. The anonymous referee is thanked for
helpful comments on the original manuscript.  GT acknowledges partial
support from NASA's Astrophysics Data Analysis Program through grant
80NSSC18K0413, and to the National Science Foundation (NSF) through
grant AST-1509375. JLC is supported by the NSF Astronomy and
Astrophysics Postdoctoral Fellowship under award AST-1602662, and by
NASA under grant NNX16AE64G issued through the \ktwo\ Guest Observer
Program (GO~7035).  This work was performed in part under contract
with the California Institute of Technology (Caltech)/Jet Propulsion
Laboratory (JPL) funded by NASA through the Sagan Fellowship Program
executed by the NASA Exoplanet Science Institute.  The research has
made use of the SIMBAD and VizieR databases, operated at the CDS,
Strasbourg, France, and of NASA's Astrophysics Data System Abstract
Service. The research was made possible through the use of the AAVSO
Photometric All-Sky Survey (APASS), funded by the Robert Martin Ayers
Sciences Fund, and made use of data products from the Wide-field
Infrared Survey Explorer, which is a joint project of the University
of California, Los Angeles, and the Jet Propulsion
Laboratory/California Institute of Technology, funded by NASA. Data
products were also used from the Two Micron All Sky Survey, which is a
joint project of the University of Massachusetts and the Infrared
Processing and Analysis Center/California Institute of Technology,
funded NASA and the NSF.  The work has also made use of data from the
European Space Agency (ESA) mission {\it Gaia}
(\url{https://www.cosmos.esa.int/gaia}), processed by the {\it Gaia}
Data Processing and Analysis Consortium (DPAC,
\url{https://www.cosmos.esa.int/web/gaia/dpac/consortium}). Funding
for the DPAC has been provided by national institutions, in particular
the institutions participating in the {\it Gaia} Multilateral
Agreement.

\end{acknowledgements}


\begin{thebibliography}

\bibitem[Andersen et al.(1991)]{Andersen:1991} Andersen, J., Clausen,
  J.\ V., Nordstr\"om, B., Tomkin, J., \& Mayor, M. 1991, \aap, 246,
  99

\bibitem[Asplund et al.(2009)]{Asplund:2009} Asplund, M., Grevesse, N.,
  Sauval, A.\ J., \& Scott, P. 2009, \araa, 47, 481 (A09)

\bibitem[Barnes(2007)]{Barnes:2007} Barnes, S.\ A.\ 2007, \apj, 669,
  1167

\bibitem[Bianchi et al.(2011)]{Bianchi:2011} Bianchi, L., Herald, J.,
  Efremova, B., et al.\ 2011, \apss, 335, 161

\bibitem[Brewer et al.(2016)]{Brewer:2016} Brewer, L.\ N., Sandquist,
  E.\ L., Mathieu, R.\ D., et al.\ 2016, \aj, 151, 66

\bibitem[Binnendijk(1960)]{Binnendijk:1960} Binnendijk, L.\ 1960,
  Properties of Double Stars, (Univ.\ of Pennsylvania Press:
  Philadelphia), p.\ 290

\bibitem[Brogaard et al.(2012)]{Brogaard:2012} Brogaard, K.,
  VandenBerg, D.\ A., Bruntt, H., et al.\ 2012, \aap, 543, A106

\bibitem[Brogaard et al.(2011)]{Brogaard:2011} Brogaard, K., Bruntt,
  H., Grundahl, F., et al.\ 2011, \aap, 525, A2

\bibitem[Buchhave et al.(2010)]{Buchhave:2010} Buchhave, L.\ A., Bakos,
  G.\ {\'A}., Hartman, J.\ D., et al.\ 2010, \apj, 720, 1118

\bibitem[Cardelli et al.(1989)]{Cardelli:1989} Cardelli, J.\ A.,
  Clayton, G.\ C., \& Mathis, J.\ S.\ 1989, \apj, 345, 245

\bibitem[Casagrande et al.(2010)]{Casagrande:2010} Casagrande, L.,
  Ram{\'{\i}}rez, I., Mel{\'e}ndez, J., Bessell, M., \& Asplund,
  M.\ 2010, \aap, 512, A54

\bibitem[Chen et al.(2014)]{Chen:2014} Chen, Y., Girardi, L., Bressan,
  A., et al.\ 2014, \mnras, 444, 2525

\bibitem[Choi et al.(2016)]{Choi:2016} Choi, J., Dotter, A., Conroy,
  C., et al.\ 2016, \apj, 823, 102

\bibitem[Claret \& Bloemen(2011)]{Claret:2011} Claret, A., \& Bloemen,
  S.\ 2011, \aap, 529, A75

\bibitem[Claret \& Cunha(1997)]{Claret:1997} Claret, A., \& Cunha,
  N.\ C.\ S.\ 1997, \aap, 318, 187

\bibitem[Curtis(2016)]{Curtis:2016} Curtis, J.\ L. 2016, PhD Thesis,
  Penn State University

\bibitem[Curtis et al.(2013)]{Curtis:2013} Curtis, J.\ L., Wolfgang,
  A., Wright, J.\ T., Brewer, J.\ M., \& Johnson, J.\ A.\ 2013, \aj, 145,
  134

\bibitem[Dotter et al.(2017)]{Dotter:2017} Dotter, A., Conroy, C.,
  Cargile, P., \& Asplund, M.\ 2017, \apj, 840, 99

\bibitem[Etzel(1981)]{Etzel:1981} Etzel, P.\ B. 1981, Photometric and
  Spectroscopic Binary Systems, Proc.\ NATO Adv.\ Study Inst.,
  ed.\ E.\ B.\ Carling \& Z.\ Kopal (Dordrecht: Reidel), p.\ 111

\bibitem[Flower(1996)]{Flower:1996} Flower, P.\ J.\ 1996, \apj, 469,
  355

\bibitem[Foreman-Mackey et al.(2013)]{Foreman-Mackey:2013}
  Foreman-Mackey, D., Hogg, D.\ W., Lang, D., \& Goodman, J.\ 2013,
  \pasp, 125, 306

\bibitem[Foreman-Mackey(2016)]{Foreman-Mackey:2016} Foreman-Mackey,
  D. 2016, The Journal of Open Source Software, 24,
  \url{http://dx.doi.org/10.5281/zenodo.45906}

\bibitem[F\H{u}r\'esz(2008)]{Furesz:2008} F\H{u}r\'esz, G. 2008, PhD
  thesis, Univ.\ Szeged, Hungary

\bibitem[Gaia Collaboration et al.(2016)]{Gaia:2016} Gaia
  Collaboration, Prusti, T., de Bruijne, J.\ H.\ J., et al.\ 2016, \aap,
  595, A1

\bibitem[Gaia Collaboration et al.(2018)]{Gaia:2018} Gaia
  Collaboration, Brown, A.\ G.\ A., Vallenari, A.\ et al.\ 2018, \aap,
  in press (arXiv:1804.09365)

\bibitem[Geller et al.(2015)]{Geller:2015} Geller, A.\ M., Latham,
  D.\ W., \& Mathieu, R.\ D.\ 2015, \aj, 150, 97

\bibitem[Gelman \& Rubin(1992)]{Gelman:1992} Gelman, A., \& Rubin,
  D.\ B. 1992, Statistical Science, 7, 457

\bibitem[Giles et al.(2017)]{Giles:2017} Giles, H.\ A.\ C., Collier
  Cameron, A., \& Haywood, R.\ D. 2017, \mnras, 472, 1618

\bibitem[Gilliland et al.(2010)]{Gilliland:2010} Gilliland, R.\ L.,
  Jenkins, J.\ M., Borucki, W.\ J., et al.\ 2010, \apjl, 713, L160

\bibitem[Goodman \& Weare(2010)]{Goodman:2010} Goodman, J., \& Weare,
  J. 2010, Commun.\ Appl.\ Math.\ Comput.\ Sci., 5, 65



\bibitem[Green et al.(2018)]{Green:2018} Green, G.\ M., Schlafly,
  E.\ F., Finkbeiner, D.\ P., et al.\ 2018, \mnras, 478, 651

\bibitem[Gregory(2005)]{Gregory:2005} Gregory, P.\ C.\ 2005, \apj, 631,
  1198

\bibitem[Grevesse \& Sauval(1998)]{Grevesse:1998} Grevesse, N., \&
  Sauval, A.\ J. 1998, Space Sci.\ Rev., 85, 161 (GS98)

\bibitem[Hartman et al.(2018)]{Hartman:2018} Hartman, J.\ D., Quinn,
  S.\ N., Bakos, G.\ {\'A}., et al.\ 2018, \aj, 155, 114

\bibitem[Henden et al.(2015)]{Henden:2015} Henden, A.\ A., Levine, S.,
  Terrell, D., \& Welch, D.\ L.\ 2015, American Astronomical Society
  Meeting Abstracts \#225, 225, 336.16

\bibitem[Henden \& Munari(2014)]{Henden:2014} Henden, A., \& Munari,
  U.\ 2014, Contributions of the Astronomical Observatory Skalnate
  Pleso, 43, 518

\bibitem[Hilditch(2001)]{Hilditch:2001} Hilditch, R.\ W.\ 2001, An
  Introduction to Close Binary Stars (Cambridge, UK: Cambridge
  University Press) p.\ 152

\bibitem[Hora et al.(1994)]{Hora:1994} Hora, J.\ L., Luppini, G.\ A.,
  \& Hodapp, K.-W. 1994, \procspie, 2198, 498

\bibitem[Huang et al.(2015)]{Huang:2015} Huang, Y., Liu, X.-W., Yuan,
  H.-B., et al.\ 2015, \mnras, 454, 2863

\bibitem[Huber et al.(2017)]{Huber:2017} Huber, D., Zinn, J.,
  Bojsen-Hansen, M., et al.\ 2017.\ 2008, \apj, 844, 102

\bibitem[Irwin et al.(2011)]{Irwin:2011} Irwin, J.\ M., Quinn, S.\ N.,
  Berta, Z.\ K., et al.\ 2011, \apj, 742, 123

\bibitem[Jenkins et al.(2010)]{Jenkins:2010} Jenkins, J.\ M., Caldwell,
  D.\ A., Chandrasekaran, H., et al.\ 2010, \apjl, 713, L87

\bibitem[Kaluzny et al.(2006)]{Kaluzny:2006} Kaluzny, J., Pych, W.,
  Rucinski, S.\ M., \& Thompson, I.\ B.\ 2006, Acta Astron., 56, 237

\bibitem[Kipping(2010)]{Kipping:2010} Kipping, D.\ M.\ 2010, \mnras,
  408, 1758

\bibitem[Kraus et al.(2016)]{Kraus:2016} Kraus, A.\ L., Ireland, M.\ J.,
  Huber, D., Mann, A.\ W., \& Dupuy, T.\ J.\ 2016, \aj, 152, 8

\bibitem[Kraus et al.(2011)]{Kraus:2011} Kraus, A.\ L., Ireland, M.\ J.,
  Martinache, F., \& Hillenbrand, L.\ A.\ 2011, \apj, 731, 8

\bibitem[Kraus et al.(2008)]{Kraus:2008} Kraus, A.\ L., Ireland, M.\ J.,
  Martinache, F., \& Lloyd, J.\ P., \apj, 679, 762

\bibitem[Latham et al.(2002)]{Latham:2002} Latham, D.\ W., Stefanik,
  R.\ P., Torres, G., et al.\ 2002, \aj, 124, 1144

\bibitem[Mandel \& Agol(2002)]{Mandel:2002} Mandel, K., \& Agol,
  E.\ 2002, \apjl, 580, L171

\bibitem[McQuillan et al.(2013)]{McQuillan:2013} McQuillan, A.,
  Aigrain, S., \& Mazeh, T. 2013, \mnras, 432, 1203

\bibitem[Meibom et al.(2015)]{Meibom:2015} Meibom, S., Barnes, S.\ A.,
  Platais, I., et al.\ 2015, \nat, 517, 589

\bibitem[Meibom et al.(2009)]{Meibom:2009} Meibom, S., Grundahl, F.,
  Clausen, J.\ V., et al.\ 2009, \aj, 137, 5086

\bibitem[Meibom \& Mathieu(2005)]{Meibom:2005} Meibom, S., \& Mathieu,
  R.\ D. 2005, \apj, 620, 970

\bibitem[Michaud(1970)]{Michaud:1970} Michaud, G.\ 1970, \apj, 160,
  641

\bibitem[Michaud et al.(1976)]{Michaud:1976} Michaud, G., Charland,
  Y., Vauclair, S., \& Vauclair, G.\ 1976, \apj, 210, 447

\bibitem[Munari \& Zwitter(1997)]{Munari:1997} Munari, U., \& Zwitter,
  T. 1997, \aap, 318, 269

\bibitem[Nordstr\"om et al.(1994)]{Nordstrom:1994} Nordstr\"om, B.,
  Latham, D.\ W., Morse, J.\ A., et al.\ 1994, \aap, 287, 338

\bibitem[Paxton et al.(2011)]{Paxton:2011} Paxton, B., Bildsten, L.,
  Dotter, A.\ et al. 2011, \apjs, 192, 3

\bibitem[Paxton et al.(2013)]{Paxton:2013} Paxton, B., Cantiello, M.,
  Arras, P., et al.\ 2013, \apjs, 208, 4

\bibitem[Paxton et al.(2015)]{Paxton:2015} Paxton, B., Marchant, P.,
  Schwab, J., et al.\ 2015, \apjs, 220, 15

\bibitem[Popper \& Etzel(1981)]{Popper:1981} Popper, D.\ M., \& Etzel,
  P.\ B. 1981, \aj, 86, 102

\bibitem[Pr{\v s}a et al.(2016)]{Prsa:2016} Pr{\v s}a, A., Harmanec,
  P., Torres, G., et al.\ 2016, \aj, 152, 41

\bibitem[Quintana et al.(2010)]{Quintana:2010} Quintana, E.\ V.,
  Jenkins, J.\ M., Clarke, B.\ D., et al.\ 2010, \procspie, 7740, 77401X

\bibitem[Rizzuto et al.(2016)]{Rizzuto:2016} Rizzuto, A.\ C., Ireland,
  M.\ J., Dupuy, T.\ J., \& Kraus, A.\ L.\ 2016, \apj, 817, 164

\bibitem[Sandquist et al.(2016)]{Sandquist:2016} Sandquist, E.\ L.,
  Jessen-Hansen, J., Shetrone, M.\ D., et al.\ 2016, \apj, 831, 11

\bibitem[Skrutskie et al.(2006)]{Skrutskie:2006} Skrutskie, M.\ F.,
  Cutri, R.\ M., Stiening, R., et al.\ 2006, \aj, 131, 1163

\bibitem[Stefanik et al.(2006)]{Stefanik:2006} Stefanik, R.\ P.,
  Latham, D.\ W., \& Davis, R.\ J.\ 2006, \pasp, 118, 1656

\bibitem[Stefanik et al.(1999)]{Stefanik:1999} Stefanik, R.\ P.,
  Latham, D.\ W., \& Torres, G.\ 1999, IAU Colloq.\ 170: Precise Stellar
  Radial Velocities, 185, 354

\bibitem[Torres et al.(2010)]{Torres:2010a} Torres, G., Andersen, J., \&
  Gim\'enez, A. 2010, \aapr, 18, 67

\bibitem[Torres(2010)]{Torres:2010b} Torres, G.\ 2010, \aj, 140, 1158 

\bibitem[Torres et al.(2002)]{Torres:2002} Torres, G., Neuh{\"a}user,
  R., \& Guenther, E.\ W.\ 2002, \aj, 123, 1701

\bibitem[Tuthill et al.(2010)]{Tuthill:2010} Tuthill, P., Lacour, S.,
  Amico, P., et al.\ 2010, \procspie, 7735, 77351O

\bibitem[Tuthill et al.(2006)]{Tuthill:2006} Tuthill, P., Lloyd, J.,
  Ireland, M., et al.\ 2006, \procspie, 6272, 62723A

\bibitem[Vanderburg \& Johnson(2014)]{Vanderburg:2014} Vanderburg, A.,
  \& Johnson, J.\ A.\ 2014, \pasp, 126, 948

\bibitem[Vanderburg et al.(2016)]{Vanderburg:2016} Vanderburg, A.,
  Latham, D.\ W., Buchhave, L.\ A., et al.\ 2016, \apjs, 222, 14

\bibitem[Voges et al.(1999)]{Voges:1999} Voges, W., Aschenbach, B.,
  Boller, T., et al.\ 1999, \aap, 349, 389

\bibitem[White et al.(2018)]{White:2018} White, T.\ R., Huber, D.,
  Mann, A.\ W.\ et al.\ 2018, \mnras, 477, 4403

\bibitem[Wright et al.(2010)]{Wright:2010} Wright, E.\ L., Eisenhardt,
  P.\ R.\ M., Mainzer, A.\ K., et al.\ 2010, \aj, 140, 1868

\bibitem[Zahn(1977)]{Zahn:1977} Zahn, J.-P.\ 1977, \aap, 57, 383 

\bibitem[Zahn(1989)]{Zahn:1989} Zahn, J.-P.\ 1989, \aap, 220, 112

\bibitem[Zucker \& Mazeh(1994)]{Zucker:1994} Zucker, S., \& Mazeh,
  T.\ 1994, \apj, 420, 806

\end{thebibliography}
\end{document}